\documentclass[CEJP,PDF]{cej} 
\usepackage{layout}
\usepackage[OT4]{fontenc}
\usepackage[latin1]{inputenc}
\usepackage{amsmath}
\usepackage{textcomp}
\usepackage{hyperref}
\usepackage{graphicx}
\usepackage{subfig}
\usepackage{float}
\title{The Lorenz model for single-mode homogeneously broadened laser: analytical determination of the unpredictible zone}

\articletype{          }

\author{Samia Ayadi\inst{1}\email{samia\_ay@yahoo.com},
        Olivier Haeberl\'{e} \inst{2}\email{olivier.haeberle@uha.fr}}

\institute{
     \inst{1} Facult\'{e} de physique, Laboratoire d'\'{e}lectronique quantique, USTHB,\\     
     Bp N 32 El Alia Bab Ezzouar ,16111 Alger,Algeria
     \inst{2} Laboratoire MIPS EA2332, Universit\'{e} de Haute-Alsace,\\
     61 rue Albert Camus, 68093 Mulhouse, France}

\abstract{We have applied harmonic expansion to derive an analytical solution for the Lorenz-Haken equations. This method is used to describe the regular and periodic self-pulsing regime of the single mode homogeneously broadened laser. These periodic solutions emerge when the ratio of the population decay rates $\wp$ is smaller than $0.11$. We have also demonstrated the tendency of the Lorenz-Haken dissipative system to behave periodic for a characteristic pumping rate "$2C_{P}$"\cite{journal-4}, close to the second laser threshold "$2C_{2th}$"(threshold of instability). When the pumping parameter ''$2C$'' increases, the laser undergoes a period-doubling sequence. This cascade of period doubling leads towards chaos. We study this type of solutions and indicate the zone of the control parameters for which the system undergoes irregular pulsing solutions. We had previously applied this analytical procedure to derive the amplitude of the first, third and the fifth order harmonics for the laser-field expansion \cite{journal-4,proceedings}. In this work, we extend this method in the aim of obtaining the higher harmonics. We show that this iterative method is indeed limited to the fifth order, and that above, the obtained analytical solution diverges from the numerical direct resolution of the equations.}


\begin{document}
\maketitle{}

\section{Introduction}
The simplest laser model is a single mode unidirectional ring laser containing a homogeneously-broadened, two-level medium, commonly designated as the Lorenz-Haken model.
In 1975, Haken \cite{journal-1} showed equivalence between the Lorenz model that describes fluid turbulence \cite{journal-2}, already known for leading to deterministic chaos, and the equations of a homogeneously broadened, single mode laser. In this context, such a laser can bee viewed as a system, which becomes unstable under suitable conditions related to the respective values of the decay rate (bad cavity condition) and of the level of excitation (second laser threshold). The numerical integration of the Lorenz-Haken model has indicated that the system undergoes a transition from a stable continuous wave output to a regular pulsing state. However, it also sometimes develops irregular pulsations (chaotic solutions). The nature of such irregular solutions was explained by Haken \cite{journal-1}.
For more than thirty years, the approach towards solving Haken-Lorenz equations was dominated by the general line of thought that ''the pulsing solutions of the single-mode laser equations must be found with numerical integration''.
In a previous work \cite{journal-3,journal-4}, it was shown that a simple harmonic expansion method permits to obtain analytical solutions for the laser equations, for physical situations where the long-term signal consists of regular pulse trains (periodic solutions). The corresponding laser field oscillates around a zero mean-value. 
In particular, we have shown that the inclusion of the third-order harmonic term in the field expansion allows for the prediction of the pulsing frequencies. The analytical expression of pulsing frequencies excellently matches their numerical counterparts, when the solution consists of regular period-one pulse trains. We have also derived a natural frequency, typical of the permanent pulsing-regime of operation.
The aim of the present work is first to extract analytical information about the zone of period doubling and about the chaotic region, from the third-order expansion analysis. Then, the  validity of the analytical development is discussed. We demonstrate in particular that the analytical expression of the periodic solutions diverges from the numerical one if one extends the development above the fifth order.

\section{Haken-Lorenz equations: period-one oscillations}
\begin{figure}[!ht]
\begin{center}
\leavevmode
\subfloat[]{%
\label{fig-temporal evolution-electric field}%
\includegraphics[height=5cm,width=7cm]{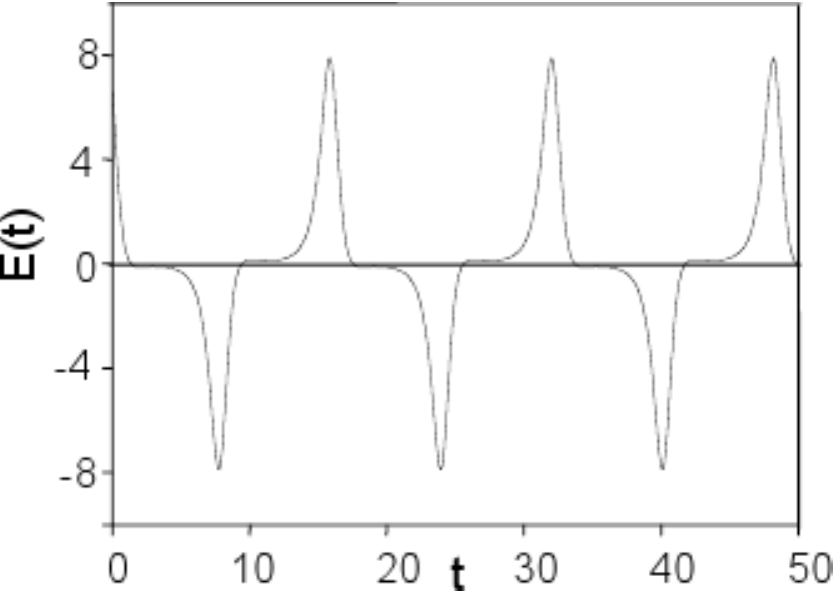}}%
 \hspace{0.9cm}
\subfloat[]{%
\label{fig-temporal evolution-FFTE}%
\includegraphics[height=5cm,width=7cm]{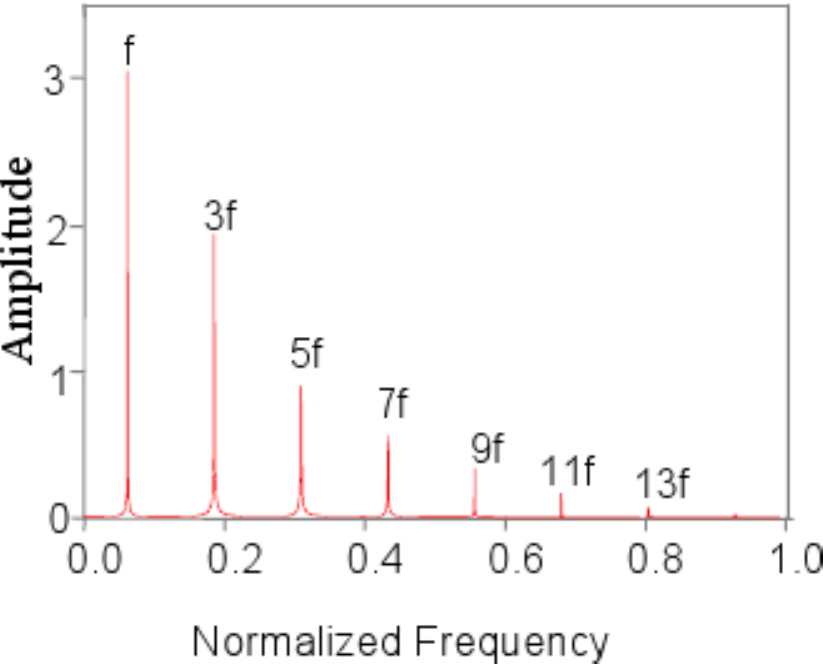}}\\
\subfloat[]{%
\label{fig-temporal evolution-polarisation}%
\includegraphics[height=5cm,width=7cm]{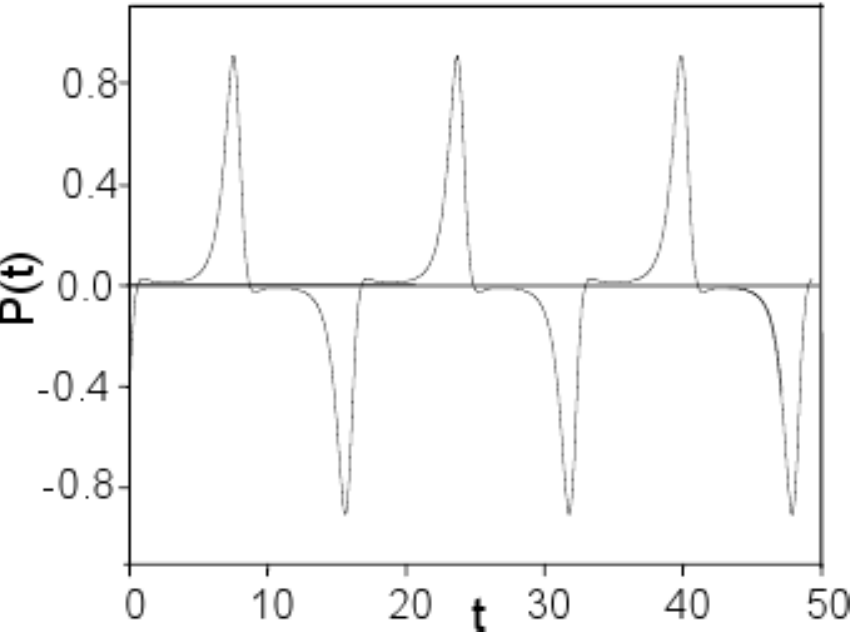}}%
\hspace{0.9cm}
\subfloat[]{%
\label{fig-temporal evolution-FFTP}%
\includegraphics[height=5cm,width=7cm]{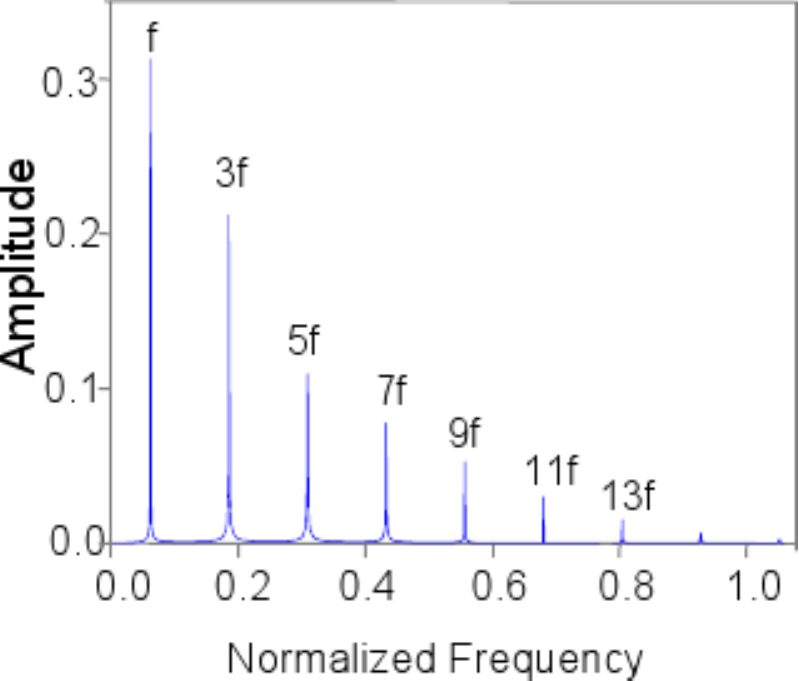}}\\
\subfloat[]{%
\label{fig-temporal evolution-population}%
\includegraphics[height=5cm,width=7cm]{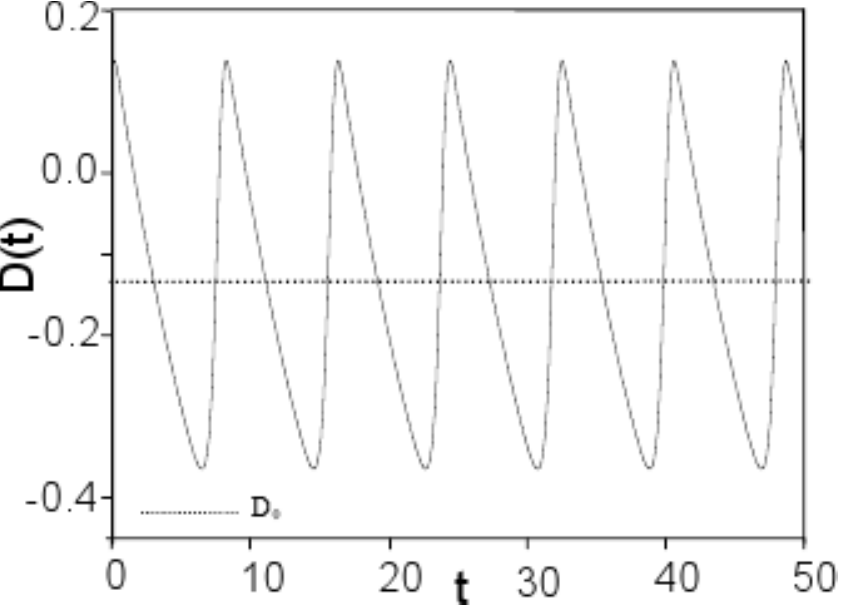}}%
\hspace{0.49cm}
\subfloat[]{%
\label{fig-temporal evolution-FFTD}%
\includegraphics[height=5cm,width=7.6cm]{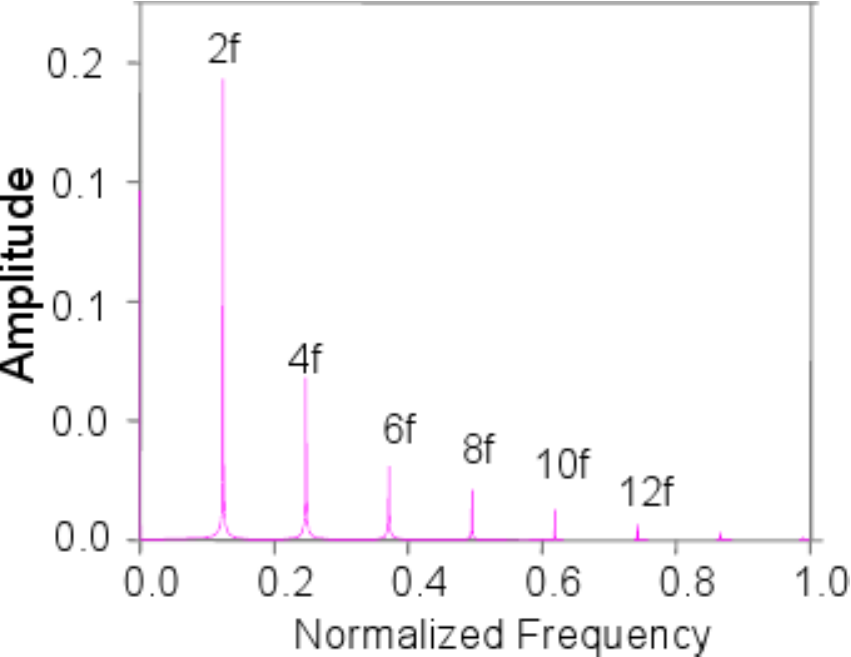}}
\caption{Long term time dependence of a) the electric-field amplitude, b) the population inversion, c) the polarization and their corresponding frequency spectra d), e) and f), respectively, for $\kappa=3$, $\wp=0.1$ and $2C=10$.}
\label{fig-temporal evolution}
\end{center}
\end{figure}

The model we start from is based on the Maxwell-Bloch equations in single mode approximation, for unidirectional ring laser containing a homogeneously broadened medium. The equations of motion are derived using a semi-classical approach, considering the resonant field inside the laser cavity as a macroscopic variable interacting with a two-level system. Assuming exact resonance between the atomic line and the cavity mode, and after adequate approximations, one obtains three differential non-linear coupled equations for the field, polarization, and population inversion of the medium, the so-called Lorenz-Haken model \cite{book-5,journal-11,journal-13}:

\begin{subequations} \label{eq1}
\begin{align}
\frac{d\,E\,\,\left( t \right)}{dt}&=-\kappa\,\left\{ {E\,\left( t\right)+2\,C\,P\,\left( t \right)} \right\}\, \label{eq1:a} \\
\frac{d\,P\,\,\left( t \right)}{dt}&=-P\,\left( t \right)+E\,\left( t\right)\,D\,\left( t \right) \label{eq1:b} \\ 
\frac{d\,D\,\,\left( t \right)}{dt}&=-\wp \,\left\{ {D\,\left( t 
\right)+1+\,P\,\left( t \right)\,\,E\,\left( t \right)} \right\}\, \label{eq1:c}
\end{align}
\end{subequations}
where $E(t)$ represents the electric field in the laser cavity having a decay constant $k$, $P(t)$ is the polarization of this field, $D(t)$ is the population difference having a decay constant $\wp$. Both $\kappa$ and $\wp$ are scaled with respect to the polarization relaxation rate, and $2C$ is the pump rate required for obtaining the lasing effect.
To obtain the steady-state solution of the equation system (\ref{eq1}), all derivatives with respect to time are set to zero. Under suitable conditions, the steady-state solution becomes unstable. We can delimit the boundary regime where Eqs. (\ref{eq1}) involve unstable solutions by linear stability analysis (LSA) \cite{book-5,book-6}. This analysis leads to the following results: the loss of stability requires simultaneously a bad cavity (i.e., $\kappa$ being sufficiently larger than $\wp+1$) and a pumping parameter $2C$ larger than $2C_{2th}$ (the instability threshold condition). This threshold $2C_{2th}$ corresponds to the onset of instability and is given by:

\begin{equation}
\label{eq2}
2C_{th} =1+\frac{\left( {\kappa+1} \right)\,\left( {\kappa+1+\wp } \right)}{\left({\kappa-1-\wp } \right)}\,
\end{equation}
At this critical value of the excitation parameter, the solution undergoes a subcritical Hopf bifurcation \cite{book-7} and loses stability, leading to a large-amplitude, pulsing solution. 
We solve numerically Eqs.(\ref{eq1}) using a Runge-Kutta method with an adaptive integration step. We represent in Fig. \ref{fig-temporal evolution} the long-term parts of the solutions obtained with the following parameter $\kappa=3$, $\wp=0.1$ and $2C=10 > 2C_{th}$ $(2C_{2th}=9.63)$. 
The temporal traces of the field Fig. \ref{fig-temporal evolution-electric field}, polarization Fig. \ref{fig-temporal evolution-polarisation} and population inversion Fig. \ref{fig-temporal evolution-population} are represented with their corresponding frequency spectra depicted on Figs. \ref{fig-temporal evolution-FFTE},\ref{fig-temporal evolution-FFTP},\ref{fig-temporal evolution-FFTD}, respectively. These spectra are obtained with a Fast-Fourier Transform algorithm. The first remark to be pointed out for this behavior is that the field and polarization oscillate around a zero mean-value, while the population inversion oscillates with a dc component (dotted line on  Fig. \ref{fig-temporal evolution-population}). The corresponding frequency spectra (Fig. \ref{fig-temporal evolution-FFTE} and Fig. \ref{fig-temporal evolution-FFTP}) exhibit odd-order components at $\Delta $, $3\Delta $, $5\Delta $,... for the field and polarization. On the contrary, the population inversion spectrum exhibits even components at $2\Delta $, $4\Delta $, $6\Delta $,...(Fig. \ref{fig-temporal evolution-FFTD}). 
These properties are at the basis of the strong-harmonic expansion method \cite{journal-3,journal-4,journal-10} that we use to construct analytical solutions.
We now give a brief outline of the main steps of the adapted strong-harmonic expansion method we use. This yields analytical expressions for the angular frequency of the periodic solutions and for the first few harmonics of the corresponding analytical solutions.\clearpage

\clearpage

\section{Iterative harmonic expansion}
We here briefly summarize the analytical procedure, which we use to determine the long-term frequencies and amplitudes of the first harmonics of the field $E(t)$ \cite{journal-3,journal-4,journal-10}. We write the interacting variables as the following expansions:
\begin{subequations} \label{eq3}
\begin{align}
E\,\left( t \right)&=\sum\limits_{n\geq0} {E_{2n+1\,\,} \cos \,\left({(2n+1)\,\Delta \,t} \right)}\, \label{eq3:a} \\
P\,\left( t \right)&=\sum\limits_{n\geq0} {P_{2n+1} \,\cos \,\left( 
{(2n+1)\,\Delta \,t} \right)} +P_{2n+2} \,\sin \,\left( {(2n+1)\,\,\Delta 
\,t} \right)\, \label{eq3:b} \\ 
D\,\left( t \right)&=D_{0} +\sum\limits_{n\geq0} {\,D_{n+1} \,\cos \,\left( 
{(2n\,)\Delta \,t} \right)} +D_{n+2} \,\sin \,\left( {(2n\,)\,\Delta \,t} 
\right)\,  \label{eq3:c}
\end{align}
\end{subequations}

Limiting these expansions to the third order for the field and polarization, and to the second order for the population inversion in Eqs. (\ref{eq3}), one obtains: 
\begin{subequations} \label{eq4}
\begin{align}
E\,\left( t \right)&=\sum\limits_{n=0,1} {E_{2n+1\,\,} \cos \,\left({(2n+1)\,\Delta \,t} \right)}\, \label{eq4:a} \\
P\,\left( t \right)&=\sum\limits_{n=0,1} {P_{2n+1} \,\cos \,\left( 
{(2n+1)\,\Delta \,t} \right)} +P_{2n+2} \,\sin \,\left( {(2n+1)\,\,\Delta\,t} \right) \label{eq4:b} \\
D\,\left( t \right)&=D_{0} +\sum\limits_{n=0,1} {D_{n+1} \,\cos \,\left({(2n\,)\Delta \,t} \right)} +D_{n+2} \,\sin \,\left( {(2n\,)\,\Delta \,t}\right)\, \label{eq4:c}
\end{align}
\end{subequations}

Inserting these truncated expansions into Eqs. (\ref{eq1}), one finds (the interested reader is referred to Refs. \cite{journal-3,journal-4,journal-10} for details of these calculations):
\begin{subequations} \label{eq5}
\begin{align}
-\Delta E_{1}&=-\kappa 2CP_{2}\, \label{eq5:a}\\ 
E_{1}&=-2CP_{1}\, \label{eq5:b} \\
-3\,\Delta E_{3}&=-\kappa 2CP_{4}\, \label{eq5:c}\\
E_{3}&=-2CP_{3}\, \label{eq5:d}
\end{align}
\end{subequations}
\begin{subequations} \label{eq6}
\begin{align}
P_{1}&=-\Gamma_{d} E_{1} +\Gamma_{d} T_{1d} \left({\wp^{2}(1-\Delta^{2})-4\,\wp \,\Delta^{2}} \right)\frac{E_{1}^{3}}{4}\, \label{eq6:a}\\
P_{2}&=-\Delta \Gamma_{d} E_{1}+2\Gamma_{d} T_{1d} \wp\Delta \left( {1+\wp -\Delta^{2}} \right)\frac{E_{1}^{3} }{4}\, \label{eq6:b}\\
P_{3}&=\,\Gamma_{d} T_{3d} \,\left\{ {\wp^{2}(1-3\Delta^{2})-8\,\wp\,\Delta^{2}} \right\}\frac{E_{1}^{3} }{4}\, \label{eq6:c}\\
P_{4}&=\,\Gamma_{d} T_{3d} 2\Delta \wp \,\left\{ {1+2\wp -3\Delta^{2}}\right\}\frac{E_{1}^{3}}{4}\, \label{eq6:d}
\end{align}
\end{subequations}
where:
\begin{align}
\Gamma_{d}&=\frac{1}{1+\Delta^{2}+\frac{E_{1}^{2}}{2}}\,\label{eq7:a}\\
T_{1d}&=\frac{1}{\left(1+\Delta^{2}\right) \left(\wp^{2}+ 4\Delta^{2}\right)}\label{eq7:b} \\
T_{3d}&=\frac{1}{\left({1+9\Delta^{2}} \right)\left({\wp^{2}+4\Delta^{2}}\right)}\label{eq7:c}
\end{align}

The expression of the angular frequency of the pulsing solution is obtained from the ratio of the first two Eqs. (\ref{eq5:a}),(\ref{eq5:b}) with the use of Eqs. (\ref{eq6:a}),(\ref{eq6:b}),(\ref{eq7:a}),(\ref{eq7:b}) and (\ref{eq7:c}). After adequate operations \cite{journal-3,journal-4}, one finds:
\begin{equation}\label{eq8}
\Delta^{2}=\frac{\left({2\,C-1} \right)\,\kappa\,\wp \,\,\left( {2+\wp}\right)-3\,\left( {\kappa+1}\right)\,\wp^{2}}{8\,\,\left( {\kappa+1} \right)-\wp\,\,\left( {2\,\kappa+\wp +4}\right)}
\end{equation}

This formula shows dependence to the pumping parameter $2C$ and indicates that the frequency of the signal increases with the excitation level. We have previously \cite{journal-3} proved that the analytical frequency expression given by Eq. (\ref{eq8}) perfectly matches the frequencies numerically derived from the equation system Eqs. (\ref{eq1}).
Another expression for the operating long-term frequency \cite{journal-4} is obtained from the ratio of the last two relations Eqs. (\ref{eq5:c}),(\ref{eq5:d}) combined with Eqs. (\ref{eq6:c}),(\ref{eq6:d}): 
\begin{equation}\label{eq9}
\frac{3\Delta}{\kappa}=-\frac{P_{4}}{P_{3}}=-\frac{2\,\Delta \,(1+2\wp-3\Delta^{2})\,\,}{\wp \,(1-3\Delta^{2})-8\Delta^{2}}\,
\end{equation}
Finally, one obtains:
\begin{equation}\label{eq10}
\Delta_{p}=\sqrt {\frac{3\,\wp +2\kappa\,\left( {1+2\wp} \right)}{24+6\kappa+9\wp}}\, 
\end{equation}
This expression shows no dependence to the excitation level $2C$. It constitutes an expression of the natural frequency that characterizes a given set of $\kappa$ and $\wp$ values that allow for periodic solutions. This frequency will be used to delimit the domain where the laser exhibits regular oscillations.

\section{Chaos via a period doubling sequence}
\begin{figure}[!ht]
\begin{center}
\leavevmode
\subfloat[]{%
\label{fig-temporals evolutions-electric field1}
\includegraphics[height=2.7cm,width=5cm]{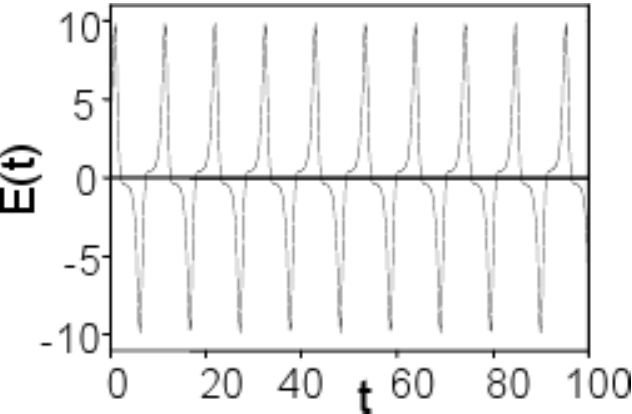}}
\hspace{0.8cm}
\subfloat[]{%
\label{fig-temporals evolutions-phase1}
\includegraphics[height=2.7cm,width=5cm]{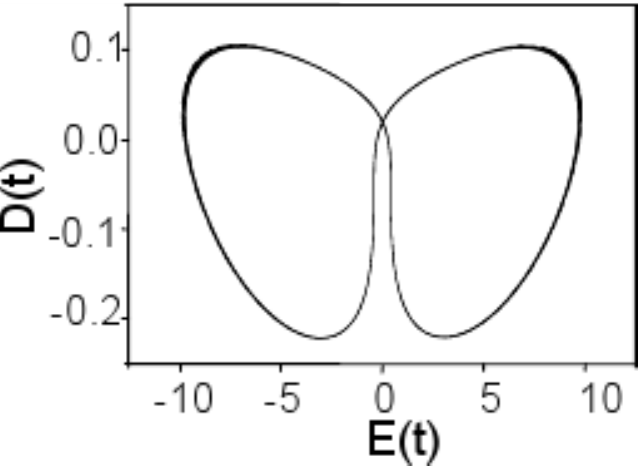}}\\
\vspace{0.3cm}\subfloat[]{%
\label{fig-temporals evolutions-electric field2}
\includegraphics[height=2.7cm,width=5cm]{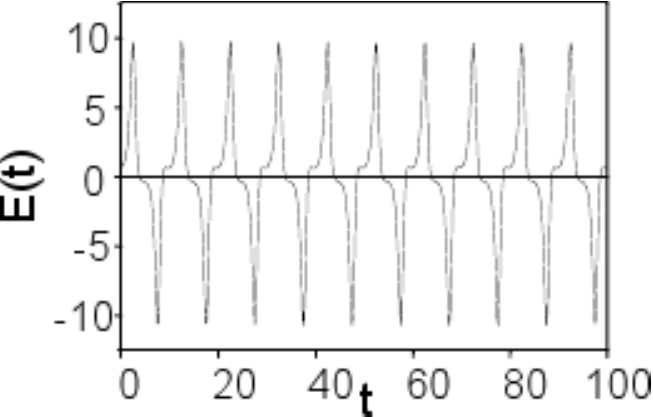}}
\hspace{0.8cm}
\subfloat[]{%
\label{fig-temporals evolutions-phase2}
\includegraphics[height=2.7cm,width=5cm]{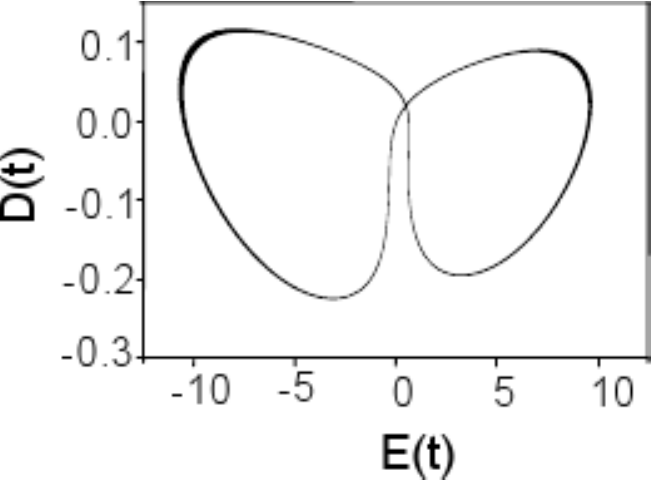}}\\
\vspace{0.3cm}\subfloat[]{%
\label{fig-temporals evolutions-electric field3}
\includegraphics[height=2.7cm,width=5cm]{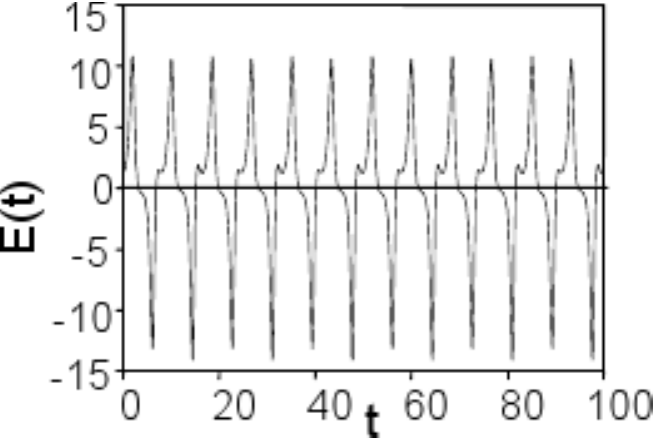}}
\hspace{0.8cm}
\subfloat[]{%
\label{fig-temporals evolutions-phase3}
\includegraphics[height=2.7cm,width=5cm]{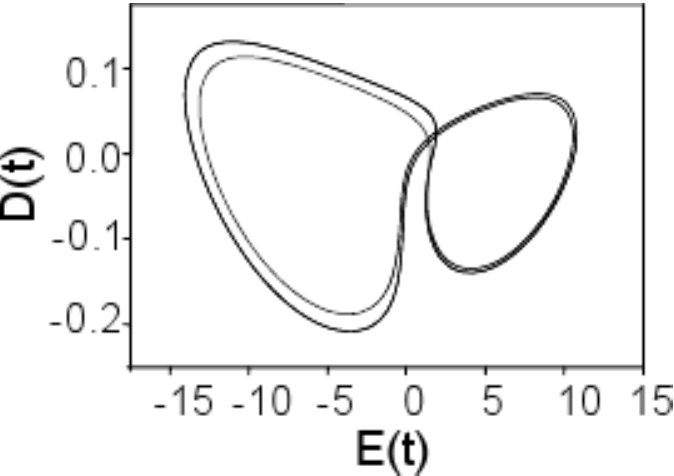}}\\
\vspace{0.3cm}\subfloat[]{%
\label{fig-temporals evolutions-electric field4}
\includegraphics[height=2.7cm,width=5cm]{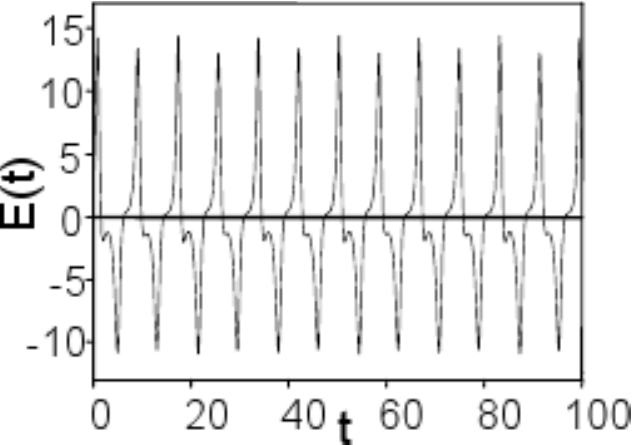}}
\hspace{0.8cm}
\subfloat[]{%
\label{fig-temporals evolutions-phase4}
\includegraphics[height=2.7cm,width=5cm]{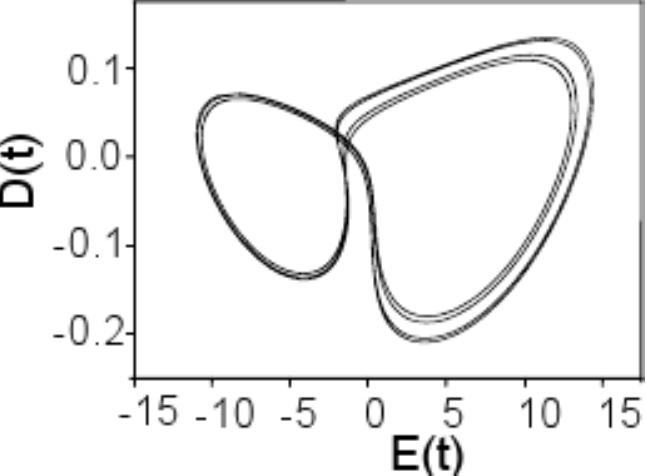}}\\
\vspace{0.3cm}\subfloat[]{%
\label{fig-temporals evolutions-electric field5}
\includegraphics[height=2.7cm,width=5cm]{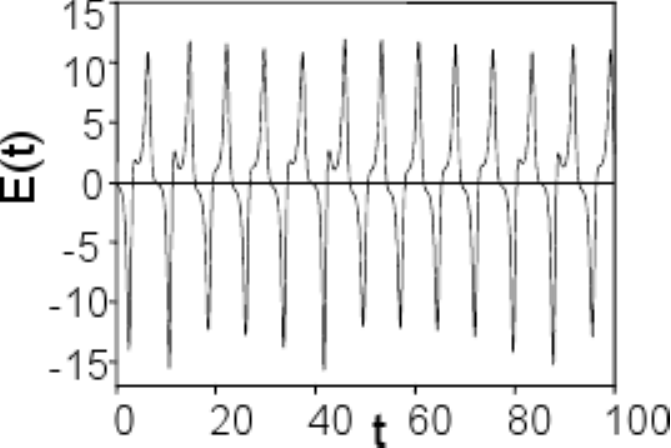}}
\hspace{0.8cm}
\subfloat[]{%
\label{fig-temporals evolutions-phase5}
\includegraphics[height=2.7cm,width=5cm]{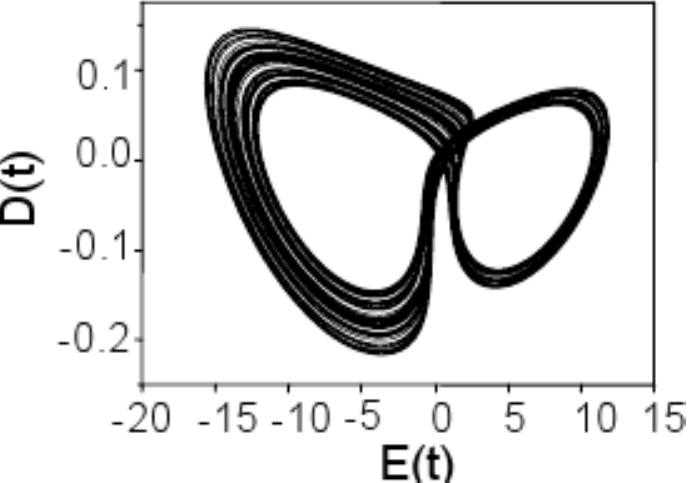}}
\caption{Examples of Eqs. (\ref{eq4}) solutions for ($\kappa=3$,$\wp=0.1$). Left column: time dependence of the electric field amplitude. a) Symmetric solution with period one, obtained with $2C=18.4$. b) Asymmetric, period-one solution, obtained with $2C=20$. c) Asymmetric solution with period two, obtained with $2C=29$. d) Asymmetric solution with period four, obtained with $2C=29.7$. e) Chaotic solution. Right column: attractor projection onto the $(D,E)$-plane. We distinguish a cascade of period doublings, which leads to chaos.}
\label{fig-temporals evolutions}
\end{center}
\end{figure}

A. Narduccsi \emph{et al}. \cite{journal-11,book-12} have shown that parameter $\wp$ plays an important role in defining the kind of dynamics predicted by the Lorenz-Haken model. Numerical simulations with $\kappa = 3$ and $\wp=0.1$ (Fig. \ref{fig-temporals evolutions}) have shown that large-amplitude periodic solutions dominate over the range $10< 2C < 32$.
The behaviour of these periodic solutions for increasing gain $2C$ is extremely variable.\\
First, the laser displays a symmetric and period-one solution (Fig. \ref{fig-temporals evolutions-electric field1}). This solution has been called symmetrical because of the symmetry of $E(t)$ with respect to $E=0$: the projection of the trajectory onto the $(E,D)$ plane produces symmetrical loops as shown in Fig. \ref{fig-temporals evolutions-phase1}.
For increasing excitation level $2C$ $(2C=18.4)$, the previously symmetric, period-one solution becomes asymmetric. The field amplitude undergoes a symmetry-breaking transformation with different positive and negative excursions (Figs. \ref{fig-temporals evolutions-electric field2},\ref{fig-temporals evolutions-phase2}). 
For $2C=29$, the laser exhibits a period-two, asymmetric solution (Figs. \ref{fig-temporals evolutions-electric field3},\ref{fig-temporals evolutions-phase3}). This feature corresponds to the period-doubling bifurcations. These results illustrates that period-doubling bifurcations can only emerge from asymmetrical solutions, as shown by Swift and Weisenfeld \cite{journal-15}.
As the excitation level further increases, more and more complicated patterns emerge, which eventually lead to chaotic behaviour. As shown in Figs. \ref{fig-temporals evolutions-electric field4},\ref{fig-temporals evolutions-phase4} and Figs. \ref{fig-temporals evolutions-electric field5},\ref{fig-temporals evolutions-phase5}, the temporal trace of the field exhibits asymmetric solution with period four when $2C=29.7$, and finally becomes chaotic for $2C=32$.\clearpage

\clearpage
We now attempt to delimit the zone where the laser can exhibit periodic pulsations using the analytical method. At first, we derive the analytical expression of the pumping rate 2C that leads the laser to oscillate with period one ($\Delta =\Delta_{p}$), period two and over. To get this expression, we assume that the pulsation frequency Eq. (\ref{eq8}) is proportional to the eigenfrequency Eq. (\ref{eq10}): 

\begin{equation}\label{eq11}
\Delta =\alpha \,\,\Delta_{p}
\end{equation}

\begin{figure}[!ht]
\begin{center}
\leavevmode
\subfloat[]{%
\label{fig-chaos-region}
\includegraphics[height=5cm,width=6.9cm]{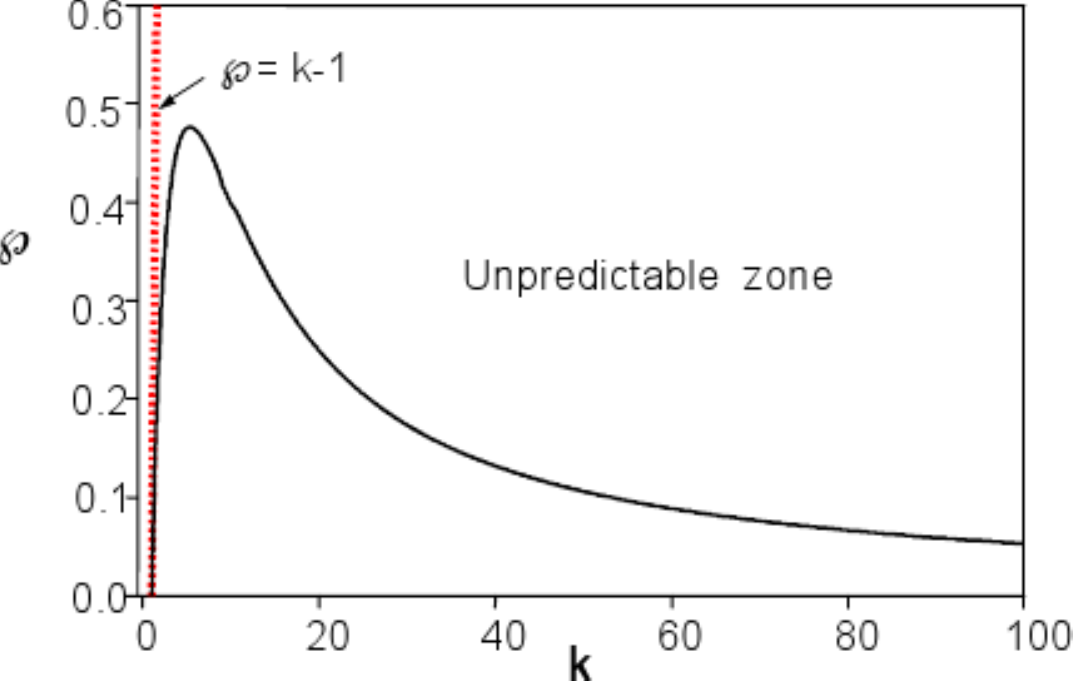}}\\
\vspace{0.3cm}\subfloat[]{%
\label{fig-chaos-Ep}
\includegraphics[height=5.2cm,width=7cm]{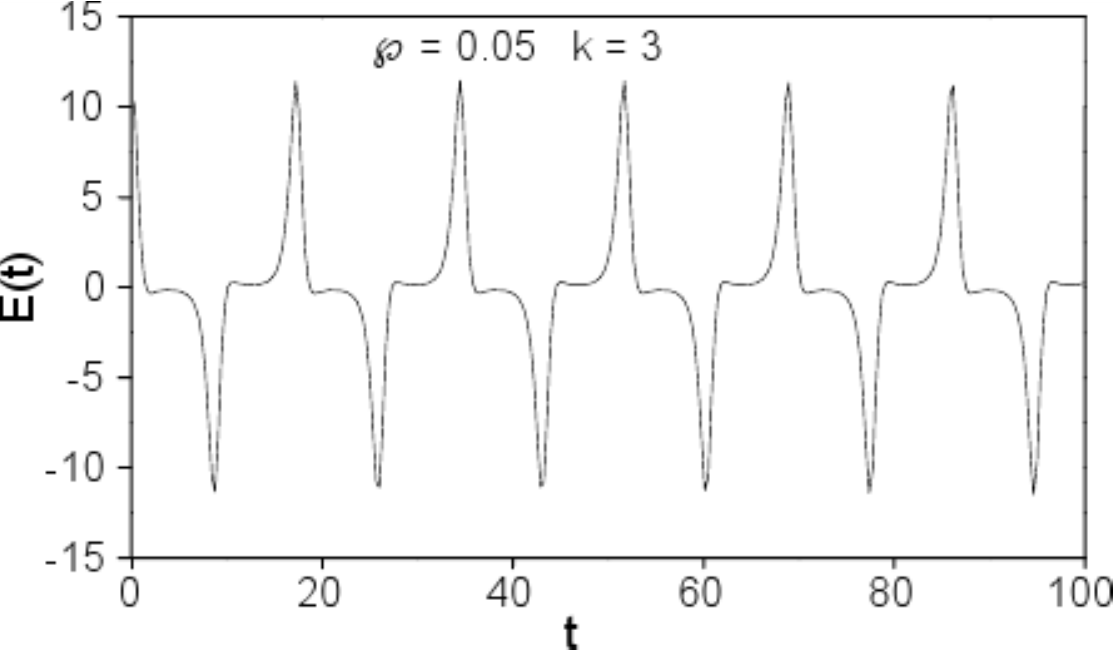}}\\
\vspace{0.3cm}\subfloat[]{%
\label{fig-chaos-Ec}
\includegraphics[height=5cm,width=7cm]{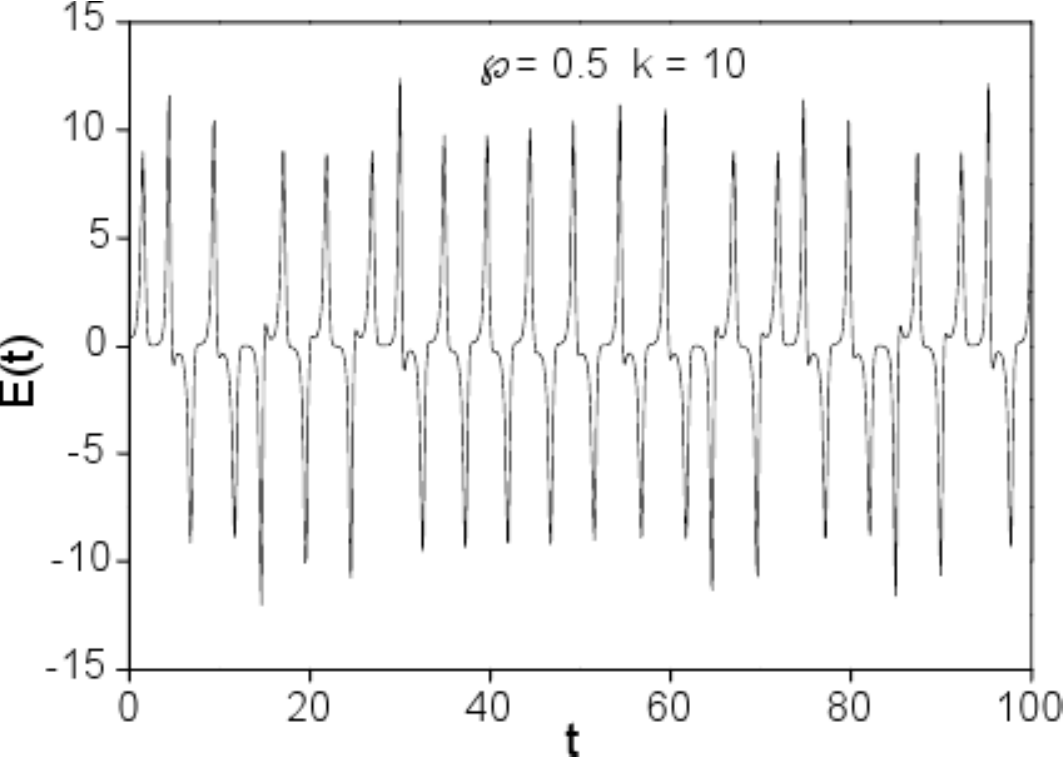}}
\caption{a) Predictable and unpredictable zone. b) Exemple of periodic solution in the region $C_{2} < C_{2th}$. c) Exemple of chaotic pulsation for $C_{2} >C_{2th}$ .}
\label{fig-chaos}
\end{center}
\end{figure}

$\alpha$ being a rational number. Then, $2C_{\alpha}$ takes the following form:
\begin{equation}\label{eq12}
2\,C_{\alpha } =\frac{(2k+3\wp +4\kappa\wp )\left\{{\alpha^{2}\left[ 
{8(\kappa+1)-\wp (2\kappa+\wp+4)} \right]+3\wp (8+2\kappa+3\wp )} \right\}}{3\kappa\wp(2+\wp 
)(8+2\kappa+3\wp)}
\end{equation}
For different values of $\alpha$, we get the pumping level $2C_{\alpha}$  that we have used for numerically solving Eqs.(\ref{eq1}). These numerical solutions for the chosen parameters $\kappa=3$ and $\wp =0.1$ can be categorized into four classes, depending on the values of $\alpha$:\\
Symmetric and period-one solutions: \hspace{6cm} $1\le \alpha <1.5$ ($9.63\le 2C\le 18.4)$\\
Asymmetric solutions with period one:\hspace{6.5cm} $\alpha \le 1.75$ ($18.4<2C\le 27.8)$\\
Asymmetric solutions with period two and over:\hspace{4.7cm} $1.75<\alpha <2$ ($27.9\le 2C<32)$\\
Chaotic solutions:\hspace{11cm} $\alpha \geq 2$ ($2C > 32)$\\  
These results are illustrated in Fig.\ref{fig-temporals evolutions}, and the route to chaos via period doubling is clearly identified. We conclude that the chaotic behaviour takes place for $\alpha\geq 2$  .
This last result enables to delimit the region of control parameters where the operation of the homogeneously-broadened single mode laser can be chaotic. To do so, we assume that $2C_{2th}$ is equal to $2C_{2}$. From this identity, we derive a characteristic equation of $\wp$ according to $\kappa$:
\begin{equation}\label{eq13}
\begin{split}
64\kappa-64\kappa^{3}+(96+208\kappa+92\kappa^{2}-40\kappa^{3}+12\kappa^{4})\wp+(120+242\kappa+274\kappa^{2}+70\kappa^{3}+6\kappa^{4})\wp^{2}\\
\quad +(24+41\kappa+76\kappa^{2}-5\kappa^{3})\wp^{3}+(15\kappa+29\kappa^{2})\wp^{4}&=0
\end{split}
\end{equation}
We have solved Eq. (\ref{eq13}) using the Mathematica{\texttrademark} software to found the analytical expressions of its roots. Accepting only solutions with physical meaning ($\wp$ positive), we obtain the curve illustrated in Fig. (\ref{fig-chaos-region}). This curve separates the periodic-solution zone (an example of periodic solution being depicted in Fig. (\ref{fig-chaos-Ep})), the so-called predictable zone, from the chaotic-operation zone (an example of chaotic solution being depicted in Fig. (\ref{fig-chaos-Ec})). This region is called the unpredictable zone, characterized by $2C_{2}>2C_{2th}$.\clearpage
\clearpage
 \section{Analytical periodic solutions}
The analytical development described in section 3 permits to establish the analytical expression of the electric field. This was done in a previous work (see appendix A in Ref. \cite{journal-4}) to derive the expressions of the first and third-order harmonics of the field. Using Eqs. (\ref{eq4:b}), (\ref{eq5:a}), (\ref{eq5:d}), (\ref{eq5:c})) and Eq. (\ref{eq7:c}), one finds:
\begin{align}
E_{1}&=-2CP_{1}=\frac{{2}\sqrt {(1+\Delta 
^{2})(4\,\Delta^{2}+\wp^{2})(1+\kappa)}}{\sqrt {\wp} \sqrt{(\wp+2\,\kappa+2\wp\,\kappa-\Delta^{2}\,(4+\wp+2\,\kappa))}}\label{eq14:a}\\
E_{3}&=-2CP_{3}=-\frac{T_{3d}(\wp^{2}(1-3\Delta^{2})-8\wp \,\Delta^{2})\frac{E_{1} 
^{3}}{4}}{1-T_{1d}(\wp^{2}(1-\Delta^{2})-4\wp \,\Delta^{2})\frac{E_{1}^{2}}{4}}\label{eq14:b}
\end{align}
The expression of $E_{1}$ and $E_{3}$ are related to the decay rates $\wp$ and $\kappa$, and to the excitation parameter $2C$ through their dependence on the pulsing frequency $\Delta$. Using the previously described analytical features, we now construct a typical sequence of analytical solutions being the counterparts of the numerical ones presented on Fig. \ref{fig-temporal evolution-electric field}-\ref{fig-temporal evolution-population}. 
The long term operating frequency is estimated from Eq. (\ref{eq8}), while the first and third order field components are evaluated from Eqs. (\ref{eq14:a}) and (\ref{eq14:b}), respectively.
The values of these components, for $\kappa=3$, $\wp=0.1$ and $2C=10$, are $E_{1}=5.19$ and $E_{3}=1.72$ and the corresponding frequency is $\Delta=0.42$. Thus, to third order, the analytical field expansion takes the following expression:
\begin{equation}\label{eq15}
E\left(t\right)=5.19\,\cos (0.42\,t)\,\,+1.72\,\cos (3\,\times\,\,0.42\,\,t)
\end{equation}
The temporal evolution of Eq. (\ref{eq15}) is illustrated in Fig. \ref{fig-divergence-E1}. One may notice the resemblance with its numerical counterpart depicted in Fig. \ref{fig-temporal evolution-electric field}, however, differences remain between the analytical and the numerical solutions. The pulses peak at $E_{n}=7.75$  in the long-term time trace of Fig. \ref{fig-temporal evolution-electric field}, while from the analytical solution expressed by Eq. (\ref{eq15}) we find $E\approx 6.95$. This difference can be attributed to the limitation of a third-order-only development. 
To verify this hypothesis, we have extended the calculations towards fifth order in field amplitude \cite{proceedings}. Hence, we expand Eqs. (\ref{eq3}) to fifth order (for $n=2$) for the electric field and polarization and adopt the same procedure as in section 3. After calculations, we obtain the fifth components in terms of the first and the third order field amplitudes:

\begin{align}
P_{5}&=\Gamma_{d} \Gamma_{5} \left\{ {f\,E_{1}^{5} +g\,E_{1}^{2} E_{3}+h\,E_{1}^{4} E_{3}+q\,E_{1} E_{3}^{2}+s\,E_{1}^{3} E_{3}^{2} } \right\}\label{eq16:a}\\
E_{5}&=-2CP_{5}=-2C\Gamma_{d} \Gamma_{5} \left\{ {f\,E_{1}^{5}+h\,E_{3}E_{1}^{4}+s\,E_{3}^{2} E_{1}^{3} +g\,E_{3} E_{1}^{2}+q\,E_{3}^{2} E_{1}}\right\}\label{eq16:b}
\end{align}
The parameters  $\Gamma_{d}$ , $\Gamma_{5}$  and the weight functions  $ f $, $ g $, $ h $, $ q $ and $ s $ , are written as:
\begin{align}
f&=-\frac{1}{16}T_{3d} \Gamma_{4} \left( {\frac{\wp \,\alpha \,_{3}}{4\Delta}-2\wp \,\Delta \alpha_{4} -5\Delta \alpha_{3}-\frac{5\thinspace\wp^{2}\Delta \alpha_{4}}{2}}\right)\label{eq17:a}\\
g&=-\frac{1}{4}\left[ {\Gamma_{4} \left( {-\frac{\wp }{4\,\Delta }+6\,\Delta+\frac{5\,\wp \,\Delta }{4\,}} \right)+\Gamma_{2} \left( {-\frac{\wp}{4\,\Delta }+6\,\Delta +\frac{5\,\wp \,\Delta }{4\,}} \right)} \right]\label{eq17:b}\\
h&=\frac{1}{16}\left[ {\begin{array}{l}
 -\Gamma_{4} T_{1d} \left( {\frac{\wp \alpha_{1} }{4\Delta }-2\wp \Delta\alpha_{2} -5\Delta \alpha_{1} -\frac{5\wp^{2}\Delta \alpha_{2} }{2}}\right) \\ 
 -\Gamma_{2} T_{1d} \alpha_{1} \left( {-5\Delta +\frac{\wp }{2\Delta }}\right)+2\Gamma_{2} T_{1d} \alpha_{2} \wp \Delta \left( {1+\frac{5\wp}{2}} \right) \\ 
 -\Gamma_{2} T_{3d} \alpha_{3} \left( {-5\Delta +\frac{\wp }{2\Delta }}\right)+2\Gamma_{2} T_{3d} \alpha_{4} \wp \Delta \left( {1+\frac{5\wp}{2}} \right) \\ 
 \end{array}} \right]\label{eq17:c}\\
q&=\Gamma_{2} \left[ {\frac{\wp }{8\,\Delta }-\Delta +5\,\Delta \,\frac{\wp}{8}} \right]\label{eq17:d}\\
s&=\frac{1}{16}\left[ {-\Gamma_{2} \,T_{1d} \left( {\frac{\alpha_{1} \,\wp}{2\,\Delta }+2\,\wp \,\Delta \,\alpha_{2} -5\,\Delta \,\left( {\alpha_{1}\,-\alpha_{2\,} \wp^{2}} \right)} \right)} \right]\label{eq17:e}
\end{align}
where:
\begin{equation}\label{eq18}
\Gamma_{2} =\frac{2\wp \,\Delta }{\wp^{2}+4\Delta^{2}},\Gamma_{4}=\frac{4\wp \,\Delta }{\wp^{2}+16\Delta^{2}},\Gamma_{5}=\frac{1}{1+25\Delta^{2}},\alpha_{1}=\wp^{2}(1-\Delta^{2}
)-4\wp \,\Delta^{2}
\end{equation}
\begin{equation}\label{eq19} 
\alpha_{2} =1+\wp -\Delta^{2},\alpha_{3} =\wp^{2}\,(1-3\Delta^{2})-8\wp \,\Delta^{2},\alpha_{4} =1+2\wp -3\Delta^{2}\
\end{equation}
For $\kappa=3$, $\wp=0.1$ and $2C=10$, we computed $E_{5}$ to get $E_{5}= 1.23$. Thus, to the fifth order, the analytical field expansion writes as:
\begin{equation}\label{eq20} 
E\left( t \right)=5.19\,\cos (0.42\,t)\,\,+1.72\,\cos (3\,\,\times\,0.42\,\,t)\,+1.23\,\cos (5\,\,\times \,0.42\,\,t)
\end{equation}
and is represented in Fig. \ref{fig-divergence-E2}. As compared to the signal of Fig. \ref{fig-temporal evolution-electric field}, the time trace of Fig. \ref{fig-divergence-E2} shows thinner and higher peak values approaching those of their numerical counterpart in Fig. \ref{fig-temporal evolution-electric field}. A priori, expanding Eqs. (\ref{eq3}) to further higher orders should lead to an even better agreement between the numerical solutions and the analytical expansion method we have adopted.
\section{Higher order for the field}
Now, we extend the calculations above the fifth order. At first, we calculate the seventh-order components, by expanding the development in Eqs. (\ref{eq3}) to $n=3$ for the field and polarization, and inserting the obtained relations in Eqs. (\ref{eq1}). The calculations are time-consuming but straightforward with the help of the Mathematica{\texttrademark} software, and one obtains the seventh-order component for the field in the form:
\begin{equation}\label{eq21} 
E_{7} =-2C[W_{7} \,E_{1}^{7}+V_{7} \,E_{1}^{6}+F_{7} \,E_{1}^{5}+H_{7}\,E_{1}^{4}+\,S_{7}\,E_{1}^{3}+R_{7} \,E_{1}^{2}+Q_{7} \,E_{1} +Z_{7}]
\end{equation}
The complete expressions of $W_{7}$, \medskip  $V_{7}$, $F_{7}$, $H_{7}$, $S_{7}$, $R_{7}$, $Q_{7}$ and $Z_{7}$ are given in Appendix A, for the interested reader who would like to make use of these developments.\\
For $\kappa=3$,$\wp=0.1$, and $2C$, one obtains:\\
\medskip $W_{7}=8.27\times 10^{-9}$, $V_{7}=1.06\times 10^{-5}$, $F_{7}=5.19\times 10^{-5}$,$H_{7}=5.21\times 10^{-4}$, $S_{7}=1.84\times 10^{-3}$,$R_{7}=-1.46\times 10^{-2}$,$Q_{7}=-5.80\times 10^{-2}$, $Z_{7}=-3.64\times 10^{-3}$.

\begin{figure}[!ht]
\begin{center}
\leavevmode
\subfloat[]{%
\label{fig-divergence-E1}
\includegraphics[height=4cm,width=7cm]{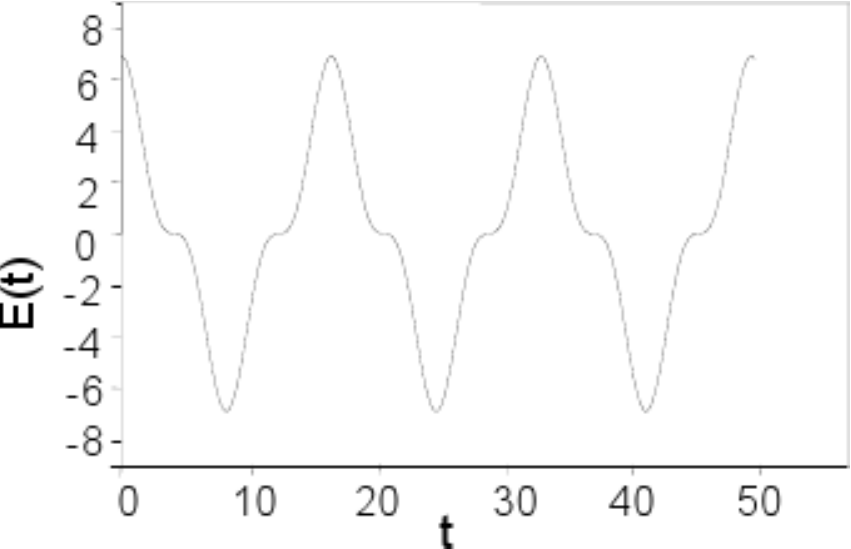}}\\
\vspace{0.2cm}\subfloat[]{%
\label{fig-divergence-E2}
\includegraphics[height=4cm,width=7cm]{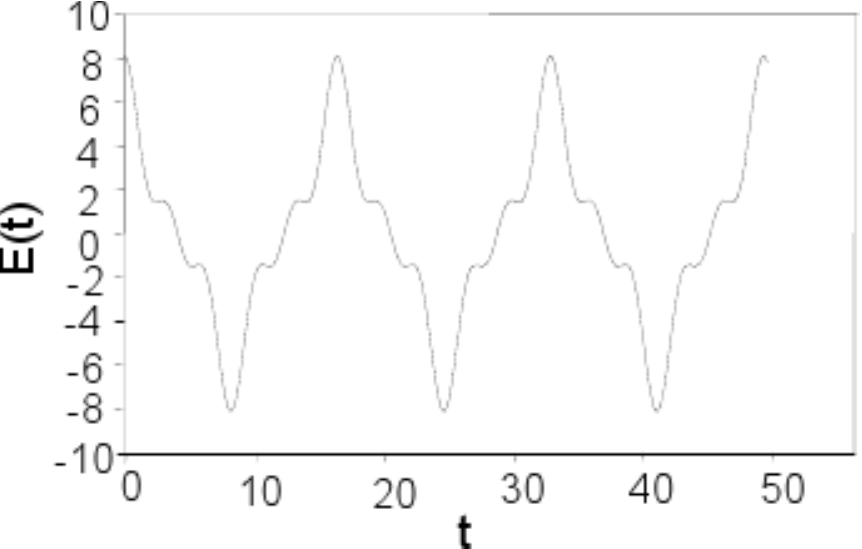}}\\
\vspace{0.2cm}\subfloat[]{%
\label{fig-divergence-E3}
\includegraphics[height=4cm,width=7cm]{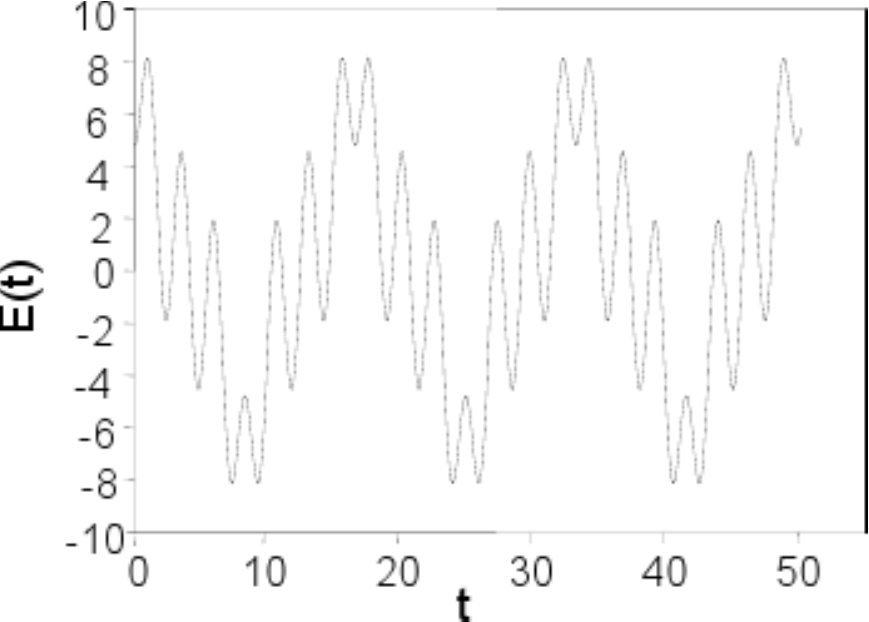}}\\
\vspace{0.2cm}\subfloat[]{%
\label{fig-divergence-E4}
\includegraphics[height=4cm,width=7cm]{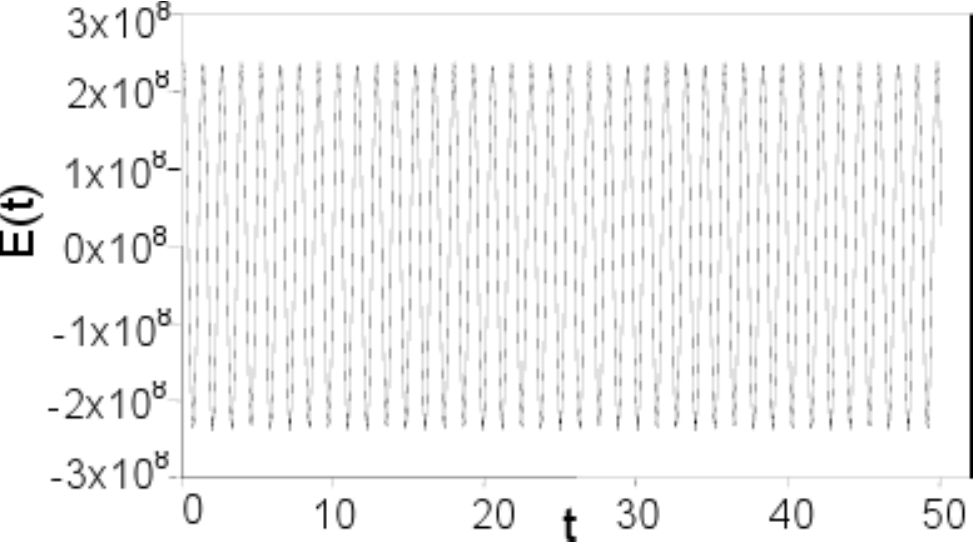}}
\caption{Analytical solutions representation for a) third-order, b) fifth-order, c) seventh-order, and d) thirteenth-order for laser-field amplitude.}
\label{fig-divergence}
\end{center}
\end{figure}

With these values, we obtain $E_{7}=-3.70$. Therefore, the analytical field expansion is given by the expression:

\begin{equation}\label{eq22}
E\left( t \right)=5.19\,\cos \left( {0.42\,t} \right)+1.73\,\cos \left( 
{3\times \,\,0.42\,t} \right)+1.23\,\cos \left( {5\,\times \,0.42\,t} 
\right)\,-3.70\,\cos \left( {7\times \,\,0.42\,t} \right)
\end{equation}
The evolution of this expression against time is shown in Fig. \ref{fig-divergence-E3}. We note that there still is a disagreement between Fig. \ref{fig-temporal evolution-electric field} and Fig. \ref{fig-divergence-E3}, owing to the negative sign of the seventh component. To correct this variance, one should take into account the higher order components in the analytical expression of the field.
Analytical evaluation of the higher terms of the field is however very lengthy and tedious, so we propose to continue the evaluation in a semi-analytical, semi-numerical way. This implies that we inject the values of parameters $\kappa$, $\wp$ and $2C$ in Eqs. (\ref{eq1}) and Eqs. (\ref{eq3}), and then evaluate the field components numerically by adopting the same procedure as previously. Figure \ref{fig-temporal evolution-FFTE} shows the field spectrum with components up to the $13^{th}$ term. We therefore compute the numerical component up to this order. For $\kappa=3$,$\wp=0.1$ and $2C=10$, such a procedure directly yields the higher-order terms:

\begin{equation} \label{eq23}
E_{9}=-64.97, \qquad E_{11}=1275.46, \qquad E_{13}=8.53\times 10^{9}
\end{equation}

The field expansion becomes:

\begin{equation}\label{eq24}
\begin{split}
E\,\left( t \right)&=5.19\,\cos \left( {0.42\,t} \right)+1.73\,\cos \left({3\,\times \,0.42\,t} \right)+1.23\,\cos \left( {5\times \,\,0.42\,t}\right)-3.34\,\cos \left( {7\,\,\times 0.42\,t} \right)\\
&\quad-64.97\cos \left( {9\,\times \,0.42\,t} \right)+12759.46\cos \left({11\,\,\times 0.42\,t} \right)+8.53\,\,10^{9}\cos \left( {13\times\,\,0.42\,t} \right)
\end{split}
\end{equation}
The time evolution of the expression given by Eq. (\ref{eq24}) is represented in Fig. \ref{fig-divergence-E4}. We remark divergence between this semi-analytical solution and the purely numerical one given on Fig. \ref{fig-temporal evolution-electric field}. This discrepancy is in fact induced by our procedure, which is iterative, and the rate of error-occurrence increases with increasing order term evaluation $(E_{11} >> E_{1}, E_{13} >> E_{1})$. This establishes the limit of this approach.

In conclusion, higher orders, above the fifth order, do not bring anymore improvement to the analytical expressions. We however believe that one could obtain a better accuracy between the analytical and the numerical solution if developing Eqs. (\ref{eq3}) up to the thirteenth order, taking into account all terms without truncation, and insert the obtained expansion in the system of equations Eqs. (\ref{eq1}). Future work will explore this approach.\clearpage
\clearpage
\section{Conclusion}
We have expanded the analytical procedure, introduced in previous work \cite{journal-3,journal-4}, that describes the self-pulsing regime of the single-mode homogeneously broadened laser operating in bad cavity configurations. The inclusion of the third-order harmonic term in the field expansion allows for deriving the analytical expression of the pumping rate $2C_{\alpha}$, that leads the laser to display period-doubling sequence. We have shown that for $\alpha < 2$, the dynamical behaviour is periodic. It is important to note that constructing analytical or semi-analytical boundaries between the various periodic regions is not an easy task, because of the non-linear behaviour of the laser. Despite these difficulties, our method allows for a satisfactory localization of the periodic solutions (predictable zone).

We have also highlighted a divergence between the numerical solution and the analytical one when we extend the evaluation above the fifth order in the expression of the laser field. Analytical expansion is not meant to replace numerical integration of the Lorenz-Haken model, which is usually the prefered procedure, but development up to the third order gives accurate results, and may be used as an alternate to direct numerical solutions, with the advantage of greater ease of use. In that case, our approach may help to adress the physics of laser chaos, especially for non-specialists.
It is interesting to mention that the Lorenz equations represent a challenge, indexed by Smale \cite{journal-8,journal-9} in its list of the fourteen problems, which constitute, according to him, a serious challenge to mathematicians. 
 \newpage
\appendix
\section{Derivation of the Seventh-Order Field Component}
We give here the weight functions of the analytical expression of the electric field $E(t)$ at seventh order $E_{7}(t)$ given by Eq. (\ref{eq21}). This expression have been derived from Eq. (\ref{eq3}), expanded up to $n = 3$, inserting these equations  into the Lorenz-Haken equations Eqs.  (\ref{eq1:a}-\ref{eq1:c}) and identifying the same order-harmonic terms in each relation. Using the Mathematica\texttrademark software, we find a hierarchical set of algebraic relations:

\begin{align}
P_{1}& = f_{1}E_{1}^{5} + h_{1}E_{1}^{4} + s_{1}E_{1}^{5} + r_{1}E_{1}^{2} + q_{1}E_{1} \\
P_{2}& = f_{2} E_{1}^{5}+ h_{2}E_{1}^{4} + s_{2}E_{1}^{5} + r_{2}E_{1}^{2} + q_{2}E_{1} \\
P_{3} &= f_{3}E_{1}^{5} + h_{3}E_{1}^{4} + s_{3}E_{1}^{5} + r_{3}E_{1}^{2} + q_{3}E_{1} + z_{3} \\
P_{4}& = f_{4}E_{1}^{5} + h_{4}E_{1}^{4} + s_{4}E_{1}^{5} + r_{4}E_{1}^{2} + q_{4}E^{1} + z_{4} \\
P_{5}& = f_{5}E_{1}^{5} + h_{5}E_{1}^{4} + s_{5}E_{1}^{5} + r_{5}E_{1}^{2} + q_{5}E_{1}\\
P_{6}& = f_{6}E_{1}^{5} + h_{6}E_{1}^{4} + s_{6}E_{1}^{5} + r_{6}E_{1}^{2} + q_{6}E_{1}\\
P_{7}& =W_{7}E_{1}^{7} + V_{7}E_{1}^{6}+ F_{7}E_{1}^{5} + H_{7}E_{1}^{4} + S_{7}E_{1}^{5} + R_{7}E_{1}^{2} + Q_{7}E_{1}+ Z_{7}
\end{align}

the seventh order field component is related to the first component through
\begin{equation}
E_{7} = - 2 C \left[ P_{7}\right]  =- 2 C \left[W_{7}E_{1}^{7} + V_{7}E_{1}^{6}+ F_{7}E_{1}^{5} + H_{7}E_{1}^{4} + S_{7}E_{1}^{5} + R_{7}E_{1}^{2} + Q_{7}E_{1}+ Z_{7}\right]
\end{equation}
with:
\begin{align}
W_{7}&=\Gamma_{10} \Gamma_{7} \left[ {\frac{\wp \,^{2}}{2}(-f_{5} +7f_{6}\Delta )+3\wp \Delta (7f_{5} \,\Delta +f_{6} \,)} \right] \\
\begin{split}
F_{7}& =\Gamma_{6} \Gamma_{7} E_{5} \left[ {\begin{array}{l} \frac{\wp \,^{2}}{2}\left( {-E_{3} f_{1} -E_{5} f_{3} -E_{3} f_{5} -h_{1}-h_{3} +7\Delta \left( {\begin{array}{l} -E_{3} f_{2} -E_{5} f_{4} +\,E_{3} f_{6} +\,h_{2} +h_{4} \end{array}} \right)} \right) \\ 
 +\wp \Delta \left( {\begin{array}{l} -E_{3} f_{2} \,-E_{5} f_{4} \,+E_{3} f_{6} +h_{2} \,+h_{4} +7\Delta \left( {E_{3} f_{1} \,+E_{5} f_{3} +E_{3} f_{5} +h_{1} \,+h_{3} }\right)  \end{array}} \right) \end{array}} \right]\\
&\quad+\Gamma_{8} \Gamma_{7} E_{3} \left[ {\begin{array}{l} \frac{\wp \,^{2}}{2}\left( {-E_{3} f_{1} -E_{5} f_{1} -h_{3} -h_{5}+7\Delta \left( {E_{3} f_{2} -E_{5} f_{2} +h_{4} +h_{6} } \right)} \right) \\ 
 +\wp \Delta \left( {\,2\left( {E_{3} f_{2} \,-E_{5} f_{2} \,+h_{4} \,+h_{6}} \right)+14\Delta \,\left( {E_{3} f_{1} \,+E_{5} f_{1} \,+h_{3} \,+h_{5} }\right)} \right) \\ 
 \end{array}} \right] \\ 
&\quad +\Gamma_{10} \Gamma_{7} \left[ {\begin{array}{l} \frac{\wp \,^{2}}{2}(-E_{5} h_{1} -3E_{3} -s_{5} +7\Delta \left( {E_{3}h_{4} +s_{6} +E_{5} \,h_{2} } \right)) \\ 
 +\wp \Delta (3E_{3} h_{4} \,+3\,s_{6} \,+3\Delta \left( {7E_{5} h_{1}+7E_{3} h_{3} \,+7\,s_{5} \wp +E_{5} h_{2} } \right)) \\  \end{array}} \right]
 \end{split}
 \end{align}
 \begin{align}
  \begin{split}
  V_{7}& =\Gamma_{6} \Gamma_{7} E_{5} \,\,\left[ {\begin{array}{l}
 \frac{\wp \,^{2}}{2}\left( {-f_{1} -f_{3} +7\Delta \left( {f_{2} +f_{4} }\right)} \right)  
 +\wp \Delta \left( {f_{2} +f_{4} \,+7\Delta \left( {f_{1} \,+f_{3} \,}\right)} \right) \\ 
 \end{array}} \right] \\ 
 &\quad+\Gamma_{10} \Gamma_{7} \left[ {\begin{array}{l} \frac{\wp \,^{2}}{2}\left( {-E_{5} f_{1} -E_{3} f_{3} -h_{5} +7\Delta\left( {E_{5} f_{2} +E_{3} f_{4} +h_{6} } \right)} \right) \\ 
 +\wp \Delta \left( {3\left( {E_{5} f_{2} \,+E_{3} f_{4} +h_{6} }\right)\,+21\Delta \left( {E_{5} f_{1} \,+E_{3} f_{3} \,+h_{5} \,} \right)}\right) \\ 
 \end{array}} \right] \\ 
&\quad+\Gamma_{8} \Gamma_{7} E_{3} \left[ {\begin{array}{l} \frac{\wp \,^{2}}{2}\left( {-f_{3} -f_{5} +7\Delta \left( {f_{4} +f_{6} }\right)} \right) \\ 
 +\wp \Delta \left( {2\left( {f_{4} \,+f_{6} } \right)\,+14\Delta \left({f_{3} \,+f_{5} \,} \right)} \right) \\ 
 \end{array}} \right]
 \end{split}\\
  \begin{split}
 Q_{7}& =\Gamma_{6} \Gamma_{7} E_{5} \left[ {\begin{array}{l} \frac{\wp^{2}}{2}\left( {-E_{3} \,q_{1} -E_{3} \,q_{5} -\,z_{3} +7\Delta\left( {-E_{3} q_{6} -E_{3} \,q_{2} +\,z_{4} } \right)} \right) \\ 
 +\wp \Delta \left( {-E_{3} \,q_{2} +E_{3} \,q_{6} +\,z_{4} +7\Delta \left({z_{3} +E_{3} \,q_{1} +E_{3} \,q_{5} } \right)} \right) \\ 
 \end{array}} \right] \\ 
&\quad+\Gamma_{8} \Gamma_{7} E_{3} \left[ {\begin{array}{l} \frac{\wp^{2}}{2}\left( {-\,z_{3} -E_{3} \,q_{1} -E_{5} \,q_{1} +7\Delta \left( {\,z_{4} +E_{3} \,q_{2} -E_{5} q_{2} } \right)} \right) \\ 
 +\wp \Delta \left( {2\left( {E_{3} q_{2} -E_{5} \,q_{2} +z_{4} }\right)+14\Delta \left( {E_{3} \,q_{1} +E_{5} \,q_{1} +\,z_{3} } \right)}\right) \\ 
 \end{array}} \right] \\ 
&\quad+\Gamma_{10} \Gamma_{7} E_{3} \left[ {\frac{\wp\,^{2}}{2}\left( {7\,z_{4} \Delta -\,z_{3} } \right)+3\wp \Delta \left({\,z_{4} +7z_{3} \Delta } \right)} \right]
\end{split}\\
 \begin{split}
  H_{7}& =\Gamma_{6} \Gamma_{7} E_{5} \left[ {\begin{array}{l} \frac{\wp \,^{2}}{2}\left( {\begin{array}{l} -E_{3} \,h_{1} -E_{5} h_{3} -E_{3} \,h_{5} -h_{1} -s_{3} + \\ 
 7\Delta \left( {-E_{3} \,h_{2} \,+E_{3} \,h_{6} \,-\,E_{5} h_{4} \,+\,E_{5}s_{2} \,+s_{4} } \right) \\ 
 \end{array}} \right) \\ 
 +\wp \Delta \left( {\begin{array}{l} \,s_{4} -E_{3} \,h_{2} -E_{5} h_{4} +E_{3} \,h_{6} +s_{2} +7\Delta \\ 
 \left( {E_{3} h_{1} +s_{1} +s_{3} +E_{5} h_{3} +E_{3} h_{5} } \right) \\ 
 \end{array}} \right) \\ 
 \end{array}} \right] \\ 
 &\quad+\Gamma_{10} \Gamma_{7} \left[ {\begin{array}{l} \frac{\wp \,^{2}}{2}\left( {-r_{5} -E_{5} s_{1} -E_{3} s_{3} +7\Delta \left( {\,E_{5} s_{2} \,+r_{6} \,+\,E_{3} s_{4} \,} \right)} \right) \\ 
 +\wp \Delta \left( {3\left( {\,r_{6} +E_{5} \,s_{2} +E_{3} s_{4} }\right)+21\Delta \left( {\,r_{5} +E_{5} s_{1} +E_{3} \,s_{3} } \right)}\right) \\ 
 \end{array}} \right] \\ 
 &\quad+\Gamma_{8} \Gamma_{7} E_{3} \left[ {\begin{array}{l}
 \frac{\wp \,^{2}}{2}\left( {-\,s_{3} -\,s_{5} -E_{3} h_{1} -\,E_{5} h_{1}+7\Delta \left( {E_{3} h_{2} \,-\,E_{5} h_{2} \,+\,s_{4} +s_{6} } \right)} \right) \\ 
 +\wp \Delta \left( {\begin{array}{l} 2\left( {E_{3} h_{2} -E_{5} h_{2} +\,s_{4} +s_{6} } \right) \\ 
 +14\Delta \left( {E_{3} h_{1} +E_{5} h_{1} +\,2s_{3} +\,2s_{5} +E_{3} h_{1}+E_{5} \,h_{1} } \right) \\ 
 \end{array}} \right) \\ 
 \end{array}} \right]  
 \end{split}\\
 \begin{split}
 S_{7}& =\Gamma_{6} \Gamma_{7} E_{5} \left[ {\begin{array}{l}
 \frac{\wp \,^{2}}{2}\left( {-r_{1} -r_{3} -E_{3\,} s_{1} -E_{5} s_{3}-E_{3} s_{5} +7\Delta \left( {r_{4} -E_{3} s_{2} -E_{5} s_{4} +E_{3} s_{6} }\right)} \right) \\ 
 +\wp \Delta \left( {\,-E_{5} s_{4} +E_{3} \,s_{6} -E_{3} \,s_{2} +\,r_{4}+7\Delta \left( {\,r_{1} +\,r_{3} +E_{3} \,s_{5} +E_{5} s_{3} } \right)}\right) \\ 
 \end{array}} \right] \\ 
 &\quad+\Gamma_{10} \Gamma_{7} \left[ {\begin{array}{l}
 \frac{\wp \,^{2}}{2}\left( {-E_{3} r_{3} -q_{5} -E_{5} r_{1} +7\Delta \left( {q_{6} +E_{3} r_{4} } \right)} \right) \\ 
 +3\wp \Delta \left( {q_{6} +E_{3} r_{4} } \right)+7\Delta \left( {q_{5}+E_{5} r_{1} +E_{3} r_{3} } \right) \\ 
 \end{array}} \right] \\ 
&\quad+\Gamma_{8} \Gamma_{7} E_{3} \left[ {\begin{array}{l}
 \frac{\wp \,^{2}}{2}\left( {-r_{5} -E_{3} s_{1} -E_{5} s_{1} -r_{3}+7\Delta \left( {r_{4} +E_{3} s_{2} -E_{5} s_{2} +r_{6} } \right)} \right)\\ 
 +\wp \Delta \left( {2\left( {E_{3} s_{2} -E_{5} s_{2} +r_{4} +\,r_{6} }\right)+14\Delta \left( {\,r_{3} +E_{5} \,s_{1} } \right)} \right) \\ 
 \end{array}} \right] 
\end{split}\\
\begin{split}
 R_{7}& =\Gamma_{6} \Gamma_{7} E_{5} \left[ {\begin{array}{l}
 \frac{\wp^{2}}{2}\left( {-q_{1} -E_{3} r_{1} -E_{5} r_{3} -E_{3} r_{5}+7\Delta \left( {q_{2} -E_{5} r_{4} +E_{3} r_{6} } \right)} \right) \\ 
 +\wp \Delta \left( {\,q_{2} -E_{5} \,r_{4} +E_{3} r_{6} +7\Delta \left({E_{3} \,r_{1} +E_{5} r_{3} +E_{3} \,r_{5} +\,q_{1} } \right)} \right) \\ 
 \end{array}} \right] \\ 
&\quad+\Gamma_{10} \Gamma_{7} E_{5} \left[ {\frac{\wp^{2}}{2}\left( {-q_{1} +7\Delta q_{2} } \right)+3\wp \Delta \left( {q_{2}+7\Delta q_{1} } \right)} \right] \\ 
&\quad+\Gamma_{8} \Gamma_{7} E_{3} \left[{\begin{array}{l}
 \frac{\wp^{2}}{2}\left( {-q_{5} -E_{3} r_{1} -E_{5} r_{1} +7\Delta q_{6} }\right) \\ 
 +\wp \Delta (2\,q_{4} +3\,q_{6} +14\Delta \left( {\,\,q_{5} +E_{3} \,r_{1}+E_{5} r_{1} } \right)) \\ 
 \end{array}} \right] 
\end{split}\\
\begin{split}
Z_{7}&=\frac{\Gamma_{6} \Gamma_{7} E_{5}^{2}}{2}\left[ {\wp \,^{2}(-z_{3}-7\,z_{4} \Delta )-2\wp \Delta (\,z_{4} +7z_{3} \Delta )} \right] \\
\end{split}
\end{align}
Where:
\begin{align}
\Gamma_{5}& =\frac{1}{\mbox{1}+25\Delta^{\mbox{2}}},\Gamma_{6} =\frac{1}{\mbox{g}^{\mbox{2}}+\mbox{4}\Delta^{\mbox{2}}},\Gamma_{8} =\frac{1}{\mbox{g}^{\mbox{2}}+16\Delta^{\mbox{2}}},\Gamma_{10} =\frac{1}{\mbox{g}^{\mbox{2}}+36\Delta^{\mbox{2}}}\\
\Gamma_{1}& =\frac{1}{1+\Delta^{\mbox{2}}},\Gamma_{3} =\frac{1}{1+9\,\Delta^{\mbox{2}}},\Gamma_{7} =\frac{1}{2(1+49\,\Delta^{\mbox{2}})}\\
 f_{5}& =\frac{\Gamma_{4} \Gamma { }_{5}T_{3d} \Gamma_{d} }{64}\left[ 
{-\frac{\wp \,\alpha_{3} }{\Delta }+20\alpha \,_{3} \Delta +2\wp 
\,^{2}\alpha \,_{4} \Delta +8\wp \,\alpha_{4} \Delta } \right]\\
\begin{split}\\ 
h_{5}& =E_{3} \Gamma_{5} \Gamma_{d} \left[ {\begin{array}{l}
 \Gamma_{2} T_{3d} \left( {\frac{\wp \,\alpha \,_{3} }{32\,\Delta 
}+\frac{5\alpha_{3} \Delta }{16}+\frac{\wp \alpha_{4} \Delta 
}{8}+\frac{5\wp \,^{2}\alpha_{4} \Delta }{16}} \right) \\ 
 +T_{1d} \left( {\begin{array}{l}
 \Gamma_{4} \left( {-\frac{\wp \alpha_{1} }{64\Delta }+\frac{5\alpha 
\,_{1} \Delta }{16}+\frac{\wp \,\alpha_{2} \Delta }{8}+\frac{5\wp 
\,^{2}\alpha \,_{2} \Delta }{32}} \right)+ \\ 
 \Gamma_{2} \left( {-\frac{\wp \,\alpha \,_{1} }{32\Delta }+\frac{5\alpha 
\,_{1} \Delta }{16}+\frac{\wp \,\alpha_{\,2} \Delta }{8}+\frac{5\wp 
\,^{2}\alpha \,_{2} \Delta }{16}} \right) \\ 
 \end{array}} \right) \\ 
 \end{array}} \right]
 \end{split}\\
 s_{5}& =E_{3}^{2}\Gamma_{2} \Gamma_{5} T_{1d} \Gamma_{d} \left[ 
{-\frac{\wp \,\alpha_{1} }{32\Delta \,}+\frac{5\,\alpha_{1} \Delta 
}{16\,}-\frac{\wp \alpha_{2} \Delta }{8}-\frac{5\,\wp \,^{2}\alpha_{2} 
\Delta }{16\,}} \right]\\
r_{5}& =E_{3} \Gamma_{5} \Gamma_{d} \left[ {\Gamma_{2} \left( {-\frac{\wp 
\,}{8\Delta \,}-\frac{3\,\Delta }{2\,}-\frac{5\,\wp \,\Delta }{8}} 
\right)+\Gamma_{4} \left( {-\frac{\wp \,}{16\Delta \,}-\frac{3\,\Delta 
}{2\,}-\frac{5\,\wp \,\Delta }{16}} \right)} \right]\\
q_{5}& =E_{3}^{2}\Gamma_{2} \Gamma_{5} \Gamma_{d} \left( {\frac{\wp 
\,}{8\Delta \,}-\Delta +\frac{5\,\wp \,\Delta }{8}} \right)\\
f_{6}& =\Gamma_{4} \Gamma_{5} T_{3d} \Gamma_{d} \left( {-\frac{\alpha_{3} 
\,}{16\,}-\frac{5\,\wp \,\alpha_{3} }{64}-\frac{\,\wp \,^{2}\alpha_{4} 
}{32}+\frac{5\,\wp \,\alpha_{4} \Delta^{2}}{8}} \right)\\
\begin{split}
h_{6}&=E_{3} \Gamma_{5} \Gamma_{d} \left( {\begin{array}{l}
 \Gamma_{2} T_{3d} \left( {-\frac{\alpha_{3} \,}{16\,}-\frac{5\,\wp 
\,\alpha_{3} }{32}-\frac{\,\wp^{2}\alpha_{4} }{16}+\frac{5\,\wp \,\alpha 
_{4} \Delta^{2}}{8}} \right)+ \\ 
 T_{1d} \left( {\begin{array}{l}
 \Gamma_{2} \left( {-\frac{\alpha_{1} \,}{16\,}-\frac{5\,\wp \,\alpha_{1} 
}{32}-\frac{\,\wp^{2}\alpha_{2} }{16}+\frac{5\,\wp \,\alpha_{2} \Delta 
^{2}}{8}} \right)+ \\ 
 \Gamma_{4} \left( {-\frac{\alpha_{1} \,}{16\,}-\frac{5\,\wp \,\alpha_{1} 
}{64}-\frac{\,\wp \,\alpha \,_{2} }{32}+\frac{5\wp \,\alpha_{2} \Delta 
^{2}}{8}} \right) \\ 
 \end{array}} \right) \\ 
 \end{array}} \right)
\end{split}\\
s_{6}& =E_{3}^{2}\Gamma_{2} \Gamma_{5} T_{1d} \Gamma_{d} \left( 
{-\frac{\alpha_{1} \,}{16\,}-\frac{5\,\wp \,\alpha_{1} }{32}+\frac{\,\wp 
^{2}\alpha_{2} }{16}-\frac{5\,\wp \,\alpha_{2} \Delta^{2}}{8}} \right)\\
r_{6}& =E_{3} \Gamma_{5} \Gamma_{d} \left( {\Gamma_{4} \left( 
{\frac{1\,}{4\,}+\frac{3\wp }{8}-\frac{5\Delta^{2}}{4}} \right)+\Gamma_{2} 
\left( {\frac{1\,}{4\,}+\frac{3\,\wp }{4}-\frac{5\Delta^{2}}{4}} \right)} 
\right)\\
q_{6}& =E_{3}^{2}\Gamma_{2} \Gamma_{5} \Gamma_{d} \left( 
{\frac{1\,}{4\,}+\frac{\wp \,}{2}+\frac{5\,\Delta^{2}}{4}} \right)\\
\begin{split}
f_{4}& =\Gamma_{d} \Gamma_{3} \left[ {\begin{array}{l} \Gamma_{2} \left( {\begin{array}{l}
 T_{1d} \left( {-\frac{\alpha_{1} }{16}-\frac{3\,\wp \,\alpha_{1} 
}{32}-\frac{\,\wp^{2}\,\alpha_{2} }{16}+\frac{3\,\wp \,\alpha_{2} \Delta 
^{2}}{8}} \right) \\ 
 +T_{3d} \left( {-\frac{\,\alpha_{3} }{16}-\frac{3\,\wp \,\alpha_{3} 
}{32}-\frac{\,\wp \,^{2}\,\alpha_{4} }{16}+\frac{3\,\wp \,\alpha_{4} 
\Delta^{2}}{8}} \right) \\ 
 \end{array}} \right) \\ 
 +\Gamma_{4} T_{3d} \left( {-\frac{\,\alpha_{3} }{16}-\frac{3\wp \,\alpha 
_{3} }{64}-\frac{\,\wp \,^{2}\,\alpha_{4} }{32}+\frac{3\,\wp \,\alpha_{4} 
\Delta^{2}}{8}} \right) \\ 
 \end{array}} \right]
\end{split}\\
\begin{split}
 h_{4}& =T_{1d} \Gamma_{d} \Gamma_{3} E_{3} \left[ {\begin{array}{l}
 \Gamma_{2} \left( {-\frac{\alpha_{1} }{16}-\frac{3\,\wp \,\alpha_{1} 
}{32}-\frac{3\,\,\wp \alpha_{2} \Delta^{2}}{8}+\frac{\,\wp^{2\,}\alpha 
_{2} }{16}} \right) \\ 
 +\Gamma_{4} \left( {-\frac{\,\,\alpha_{1} }{16}-\frac{\,3\wp \alpha_{1} 
}{64}-\frac{\,\wp^{2}\alpha_{2} }{32}-\frac{3\,\,\wp \,\alpha_{2} \Delta 
^{2}}{8}} \right)-\frac{3\,\alpha_{1} \Delta }{8} \\ 
 \end{array}} \right]
\end{split}\\
 s_{4}& =\Gamma_{3} \Gamma_{d} \left[ {\Gamma_{2} \left( 
{\frac{1}{4}+\frac{\wp \,}{2}-\frac{3\Delta^{2}}{4}} \right)+-\frac{3E_{3} 
^{2}T_{3d} \alpha_{3} \Delta }{8}} \right]\\
r_{4}& =\Gamma_{3} \Gamma_{d} E_{3} \left[ {\frac{\Gamma_{3} \Gamma_{d} 
}{4}\left( {1+\wp +3\Delta^{2}} \right)+\left( {\frac{\Gamma_{4} 
}{4}\left( {1+\wp -3\Delta^{2}} \right)+\frac{3\Delta }{2}} \right)} 
\right]\\
z_{4}&=-3E_{3} \Gamma_{3} \Delta
\end{align} 
\begin{align}
\begin{split}
f_{3}&=\Gamma_{3} \Gamma_{d} \left[ {\begin{array}{l}
 \Gamma_{2} \left( {\begin{array}{l}
 T_{1d} \left( {-\frac{\,\wp \alpha_{1} }{32\Delta }+\frac{\,3\,\,\alpha 
_{1} \Delta }{16\Delta }+\frac{\,\wp \,\,\alpha_{2} \Delta 
}{8}+\frac{3\,\wp^{2}\alpha_{2} \Delta }{16}} \right) \\ 
 +T_{3d} \left( {-\frac{\,\wp \,\,\alpha_{3} }{32\Delta }+\frac{\,\wp 
\alpha_{4} \Delta }{8}+\frac{\,3\wp^{2}\alpha_{4} \Delta 
}{16}+\frac{\,3\,\,\alpha_{3} \Delta }{16}} \right) \\ 
 \end{array}} \right) \\ 
 +\Gamma_{4} T_{3d} \left( {\frac{\,3\,\,\alpha_{3} \Delta 
}{16}-\frac{\,\wp \,\,\alpha_{3} }{64\Delta }+\frac{\,\wp \,\alpha_{4} 
}{8}+\frac{\,3\wp^{2}\alpha_{4} \Delta }{32}} \right) \\ 
 \end{array}} \right]
 \end{split}\\
\begin{split}
h_{3}&=\Gamma_{3} \Gamma_{d} T_{1d} E_{3} \left[ {\begin{array}{l}
 \Gamma_{2} \left( {-\frac{\,\,\wp \,\alpha_{1} }{32\Delta 
}+\frac{\,3\,\alpha_{1} \Delta }{16}-\frac{\,\,\wp \,\alpha_{2} \Delta 
}{8}-\frac{\,3\,\wp^{2}\,\alpha_{2} \Delta }{16}} \right) \\ 
 +\Gamma_{4} \left( {\frac{\,\,\wp \,\alpha_{2} \Delta }{8}+\frac{\,3\,\wp 
^{2}\,\alpha_{2} \Delta }{32}-\frac{\,\,\wp \,\alpha_{1} }{64\Delta 
}+\frac{\,3\,\alpha_{1} \Delta }{16}} \right)-\frac{\,\,\alpha_{1} }{8} \\ 
 \end{array}} \right]
\end{split}\\
s_{3}&=\,\Gamma_{3} \Gamma_{d} \left[ {-\frac{\,E_{3}^{2}\,T_{3d} 
\,\alpha_{3} }{8}+\Gamma_{2} \left( {\frac{\wp \,\,}{8\Delta }-\frac{3\wp 
\,\Delta }{8}-\Delta } \right)} \right]\\
r_{3}& =\Gamma_{3} \Gamma_{d} E_{3} \left[ {\frac{1\,\,}{2}+\Gamma_{2} 
\left( {\frac{\,\,\wp \,\,}{8\Delta }-\frac{\,\,\Delta }{2}+\frac{\,3\,\wp 
\,\Delta }{8}} \right)+\Gamma_{4} \,\left( {-\frac{\,3\Delta \wp 
}{16}+\frac{\,\,\wp }{16\Delta }-\Delta } \right)} \right]\\
z_{3}& =-E_{3} \Gamma_{3}\\
\begin{split}
f_{2} =\Gamma_{1} \Gamma_{d} \left[ {\,\Gamma_{2} \left( 
{\begin{array}{l}
 T_{1d} \left( {-\frac{\,\alpha_{1} }{16}-\frac{\,\wp \,\alpha_{1} 
}{32}+\frac{\,\wp \,\alpha_{2} \Delta^{2}}{8}-\frac{\,\wp^{2}\alpha_{2} 
}{16}} \right) \\ 
 +\,T_{3d} \left( {-\frac{\,\alpha_{3} }{16}-\frac{\,\wp \,\alpha_{3} 
}{32}-\frac{\,\wp^{2}\alpha_{4} }{16}+\frac{\,\wp \,\alpha_{4} \Delta 
^{2}}{8}} \right) \\ 
 \end{array}} \right)-\frac{\,\,T_{1d} \alpha_{1} \Delta }{8}} \right]
\end{split}\\
\begin{split}
 h_{2}& =\Gamma_{1} \Gamma_{d} E_{3} \left[ {\begin{array}{l}
 \Gamma_{2} \left( {\begin{array}{l}
 T_{1d} \left( {-\frac{\,\wp \,\,\alpha_{1} }{16}+\frac{\,\wp^{2}\,\alpha 
_{2} }{8}} \right)+ \\ 
 T_{3d} \left( {\frac{\,\,\alpha_{3} }{16}-\frac{\,\wp \,\alpha_{3} 
}{32}+\frac{\,\wp^{2}\,\alpha_{4} }{16}+\frac{\,\wp \,\,\alpha_{4} \Delta 
^{2}}{8}} \right) \\ 
 \end{array}} \right) \\ 
 +T_{3d} \left( {\Gamma_{4} \left( {-\frac{\,\,\alpha_{3} 
}{16}-\frac{\,\wp \,\,\alpha_{3} }{64}-\frac{\,\wp^{2}\,\alpha_{4} 
}{32}+\frac{\,\wp \,\alpha_{4} \Delta^{2}}{8}} \right)-\frac{\,\,\,\Gamma 
_{3} \alpha_{3} \Delta }{8}} \right) \\ 
 \end{array}} \right]
\end{split}\\
 \begin{split}
s_{2}& =\Gamma_{1} \Gamma_{d} \left[ {\begin{array}{l} \Gamma_{2} \left( {\frac{\,1\,}{4}+\frac{\,\,\gamma \,}{4}+T_{1d} \left({\begin{array}{l} E_{3}^{2}\left( {\frac{\,\,\alpha_{1} }{16}-\frac{\,\,\wp \,\alpha_{1} 
}{32}-\frac{\,\,\wp^{2}\alpha_{2} }{16}-\frac{\,\wp \,\alpha_{2} \Delta^{2}}{8}} \right) \\ 
 -\frac{\Delta^{2}}{4} \\ 
 \end{array}} \right)} \right) \\ 
 +\Gamma_{4} T_{1d} E_{3}^{2}\left( {-\frac{\,\,\alpha_{1}}{16}-\frac{\,\,\wp \,\alpha_{1} }{64}-\frac{\,\,\wp^{2\,}\alpha_{2}}{32}+\frac{\,\wp \,\alpha_{2} \Delta^{2}}{8}} \right)+\frac{\,\Delta }{2}\\ 
 \end{array}} \right]
\end{split}\\
 q_{2}& =\Gamma_{1} \left[ {\Gamma_{d} E_{3}^{2}\left( {\Gamma_{2} \left( 
{-\frac{\,1\,}{4}+\frac{\,\wp \,}{4}-\frac{\,\Delta^{2}}{4}} \right)+\Gamma 
_{4} \left( {\frac{\,1\,}{4}+\frac{\,\wp \,}{8}-\frac{\,\Delta^{2}}{4}} 
\right)} \right)-\Delta } \right]\\
\begin{split}
f_{1}&=\Gamma_{1} \Gamma_{d} \left[ {-\frac{\,T_{1d} \alpha_{1} 
}{8}+\Gamma_{2} \left( {\begin{array}{l}
 T_{1d} \left( {-\frac{\,\,\wp \,\alpha_{1} }{32\Delta }+\frac{\,\,\alpha 
_{1} \Delta }{16}+\frac{\,\,\wp \,\alpha_{2} \Delta }{8}+\frac{\,\wp 
\,^{2}\,\alpha_{2} \Delta }{16}} \right) \\ 
 +T_{3d} \left( {\frac{\,\,\alpha_{3} \Delta }{16}+\frac{\,\wp \,\alpha 
_{4} \Delta }{8}+\frac{\,\wp^{2\,}\alpha_{4} \Delta }{16}-\frac{\,\,\wp 
\,\alpha_{3} }{32\Delta }} \right) \\ 
 \end{array}} \right)} \right]
\end{split}\\
\begin{split}
h_{1}& =\Gamma_{1} \Gamma_{d} E_{3} \left[ {\begin{array}{l}
 \Gamma_{2} \left( {\begin{array}{l} T_{1d} \left( {-\frac{\,\wp \,\,\alpha_{1} }{16\Delta }-\frac{\,\wp 
^{2}\,\alpha_{2} \Delta }{8}} \right)+ \\ 
 T_{3d} \left( {-\frac{\,\wp \,\,\alpha_{3} }{32\Delta }-\frac{\,\,\alpha_{3} \Delta }{16}+\frac{\,\wp \,\alpha_{4} \Delta }{8}-\frac{\,\wp^{2}\,\alpha_{4} \Delta }{16}} \right) \\ 
 \end{array}} \right) \\ 
 +\Gamma_{4} T_{3d} \left( {\frac{\,\wp^{2}\,\alpha_{4} \Delta 
}{32}-\frac{\,\wp \,\,\alpha_{3} }{64\Delta }+\frac{\,\,\alpha_{3} \Delta 
}{16}+\frac{\,\wp \,\alpha_{4} \Delta }{8}} \right)-\frac{\,\,\alpha_{3} 
T_{3d} }{8} \\ 
 \end{array}} \right]
\end{split}\\
\begin{split}
s_{1}& =\Gamma_{1} \Gamma_{d} \left[ {\begin{array}{l}
 \Gamma_{2} \left( {\begin{array}{l} \frac{\,\wp \,\,}{8\Delta }-\frac{\,\,\Delta }{2}-\frac{\,\wp \,\Delta 
}{8}+ \\  T_{1d} E_{3}^{2}\left( {-\frac{\,\wp \alpha_{1} }{32\Delta 
}-\frac{\,\alpha_{1} \Delta }{16}-\frac{\,\wp \,\,\alpha_{2} \Delta 
}{8}+\frac{\,\wp^{2}\alpha_{2} \Delta }{16}} \right) \\ 
 \end{array}} \right) \\ 
 +\Gamma_{4} T_{1d} E_{3}^{2}\left( {\frac{\,\wp \,\,\alpha_{2} \Delta 
}{8}+\frac{\,\wp^{2}\,\alpha_{2} \Delta }{32}-\frac{\,\wp \,\,\alpha_{1} 
}{16\Delta }+\frac{\,\,\alpha_{1} \Delta }{16}} \right)+\frac{1}{2} \\ 
 \end{array}} \right]\\
\end{split}\\
r_{1}& =\frac{\Gamma_{1} \Gamma_{2} \Gamma_{d} E_{3} }{4}\wp \left[ 
{\frac{\,1\,}{\Delta }+\,\,\Delta } \right]\\
q_{1}& =\Gamma_{1} \,\left[ {-1+\Gamma_{d} E_{3}^{2}\left( {\Gamma_{2} 
\left( {\frac{\,\wp \,}{8\Delta }+\frac{\,\Delta }{2}-\frac{\,\wp \Delta 
}{8}} \right)+\Gamma_{4} \,\left( {\frac{\wp \,\,}{16\Delta 
}-\frac{\,\Delta }{2}-\frac{\,\wp \,\Delta }{16}} \right)} \right)} \right]
\end{align} 
\end{document}